\documentclass[prc,preprint]{revtex4}
\usepackage{graphicx,latexsym}
\usepackage{dcolumn}

\begin{document}

\title{Ergodicity of the $\Delta_3$ statistic and purity of neutron resonance data}

\author{Declan Mulhall}
 \affiliation{Department of Physics/Engineering,
 University of Scranton, Scranton, Pennsylvania 18510-4642, USA.}
 \email{mulhalld2@scranton.edu}
\author{Zachary Huard}
 \affiliation{Department of Physics, University of Cincinnati,
 400 Geology/Physics Building P.O. Box 210011
Cincinnati, Ohio 45221-0011, USA.}
\author{Vladimir Zelevinsky }
 \affiliation{Department of Physics and Astronomy and National
 Superconducting Cyclotron Laboratory, Michigan State University,
 East Lansing , Michigan 48824-1321, USA .}

\date{\today}

\begin{abstract}
The $\Delta_3(L)$ statistic characterizes the fluctuations of
the number of levels as a function of the length of the
spectral interval. It is studied as a possible tool to
indicate the regular or chaotic nature of the underlying
dynamics, to detect missing levels and the mixing of
sequences of levels of different symmetry, particularly in
neutron resonance data. The relation between the ensemble
average and the average over different fragments of a given
realization of spectra is considered. A useful expression for
the variance of $\Delta_3(L)$ which accounts for finite
sample size is discussed. An analysis of neutron resonance
data presents the results consistent with a maximum
likelihood method applied to the level spacing distribution.
\end{abstract}


\maketitle

\section{\label{sec:level1}Introduction}

Neutron resonances provided the first context for modeling physical
reality with Random Matrix Theory (RMT) \cite{porterbook}; for a
brief history of RMT see \cite{guhr}. Bohr's compound nucleus
description \cite{bohr} identified the positions of the resonances
as eigenvalues of the unknown complicated Hamiltonian governing the
compound nucleus. Calculating the energies of these excited states
is impossible, even for non-interacting particles the level density
is prohibitive just from combinatorics alone; with interactions
included, exact calculations even in a reasonably truncated space
quickly become impossible. However, robust statistical features of
the spectra are calculable.  Statistical spectroscopy had already
been opened up by Gurevich in 1939 when he investigated the
regularities of level spacings in nuclear spectra \cite{gurevich}.
Wigner took it a step farther in 1951 when he suggested that
although the specific energies cannot be calculated, the wave
functions are so complicated that the statistical behavior of the
energy levels mimics that of the eigenvalues of an ensemble of
random matrices whose elements have probability distributions that
do not favor any particular basis, i.e. their statistics are
invariant under orthogonal transformations (change of basis). The
underlying assumption is that the actual Hamiltonian is complicated
in any basis excluding some exceptional ones which form a manifold
of measure zero. This led, for a time-reversal invariant physical
system, to the Gaussian Orthogonal Ensemble (GOE) and, for different
global symmetries, to other ``canonical" ensembles.

In a sense, canonical Gaussian ensembles describe the extreme
degree of universal chaoticity, and the real interest is in
understanding when, and to what extent, this limit is
realized in actual physical systems. The spectra of quantum
systems whose classical counterparts are chaotic are well
modeled by the GOE: they have the same spectral fluctuation
properties \cite{brody, stock, boh84}. This leads us to
accept RMT as a working definition of quantum chaos: a
quantum system is deemed chaotic if its spectra exhibit the
same local fluctuation properties as those of the appropriate
Gaussian ensemble. This definition frees us from the need to
work backwards from classical mechanics.

The correspondence between the GOE and nuclear spectra has
been verified many times, over a surprising range of
energies. Careful analysis showed that RMT agreed well with
the neutron \cite{liou72, liou75, jain,frankle94} and proton
\cite{watson81} resonance data for various isotopes. The
range of energies at which the nucleus exhibited signatures
of chaos was extended all the way to the ground state region
in odd-odd nuclei and to two-quasiparticle threshold in
even-even nuclei \cite{abul85, shriner91, raman91, abul96}.
Furthermore, shell model calculations exhibit many of the
fluctuation properties of the GOE \cite{big,horoiran}; for an
account of tests of RMT in nuclei see \cite{mitch01}. Ongoing
experimental high-precision studies of neutron resonances in
heavy nuclei \cite{koehler07} require  more attention to the
details and statistical justification of RMT in its practical
applications.

Thus RMT provides us with an arsenal of tools, in the form of
certain useful statistics, to analyze neutron resonance data
\cite{dyson}, the largest available body of nuclear spectra. A
statistic is a number, $W$, which can be computed from a sequence of
levels, and whose mean, $\langle W \rangle$, and variance, ${\rm
Var(W)} = \langle W^2\rangle-\langle W \rangle^2$, are calculable
from theory, and which has a small deviation from the theoretical
value when the theoretical model is a valid one. In our case, the
theoretical model is that the local fluctuations of the energy
levels of excited nuclei are described by RMT. To use these
statistics, the experimental energies must be rescaled to the
uniform level density. This process, called unfolding, removes
secular variations of the spectra, leaving just the fluctuations
about the mean values, thus allowing spectra from different physical
systems to be compared.

Among various statistical measures of spectral fluctuations,
the neighboring level spacing distribution $P(s)$, and the
Dyson-Mehta $\Delta_{3}(L)$ statistic that quantifies the
fluctuations of the number of levels in a given spectral
interval, are the most useful in practice. Even their
qualitative features allow one to quickly get a first glimpse
of the character of underlying dynamics, $-$ regular, chaotic
or intermediate. The quantitative analysis provides more
detailed characteristics. In order to apply such measures
successfully, one needs to know the completeness of a
fragment of an empirical spectrum and its purity. Missing
levels or contamination by levels of different symmetry
classes leads to distortions of the statistics. Apart from
that, an important question is that of the {\sl ergodicity}
of those statistical measures. The predictions of RMT refer,
as a rule, to  the ensemble average of the quantity of
interest. Experiment, on the other hand, typically gives the
spectral average of the quantity inside an observed subset of
the large, formally infinite, spectrum, which is taken as a
random representative of an ensemble. The ergodic property
means that the same results are valid for different fragments
of a given spectrum as well as for the average over the
ensemble.

The Dyson-Mehta $\Delta_3(L)$ statistic will be the focus of
this work. We study fluctuations of this statistic, its
ergodic properties and sensitivity to missing levels and
impurities. The point of contact with experimental data will
be neutron resonance data. There are many other systems that
lend themselves to an RMT analysis. The sizes of typical data
sets vary from system to system. The spectrum of
electromechanical vibrations in a quartz block \cite{guhr},
for example, can have many hundreds of levels, as can data
from superconducting microwave cavities. Many of the
calculations in this paper have $N$ in the hundreds. The main
results are still applicable to neutron resonance data, even
though the majority of experimental data have less than 100
levels. We provide an error estimate on the calculation of
the fraction of missed levels for data sets with $N=100$.

In Sec.  \ref{sec:d3def} we define the $\Delta_3(L)$ and discuss its
ensemble average. In Sec. \ref{sec:d3disc} we discuss spectral and
ensemble averaging, the ergodicity of the $\Delta_3(L)$ statistic,
and the corresponding uncertainties. In Sec. \ref{sec:d3} the
calculation of GOE spectra, $\Delta_3(L)$, and the unfolding
procedure will be described. $\Delta_3(L)$ is calculated exactly,
with no numerical minimization procedures. Following that Sec.
\ref{sec:sig} deals with the calculation of the uncertainties. An
analysis of actual neutron resonance data with the maximum
likelihood method is described in Sec. \ref{sec:ps}. In Sect.
\ref{sec:data} we perform a $\Delta_3(L)$ analysis of neutron
resonance data and compare it with the maximum likelihood method. We
summarize all our findings in the Conclusion.

\section{\label{sec:d3def}$\Delta_{3}(L)$ statistic: definition and ensemble average}

By definition,
\begin{eqnarray}
\Delta_{3}(L) &=\left \langle {\rm min}_{A,B}\;
\frac{1}{L}\;\int^{E_i+L}_{E_i}dE'\,[\;{\mathcal
N}(E')-AE'-B]^{2}\;
\;\right\rangle \\
&=\langle \Delta^i_3(L) \rangle\:, \label{eq:d3}
\end{eqnarray}
where we use the notation $\langle x \rangle$ for the {\sl
spectral average} of $x$. This is a measure of the average
deviation of the spectrum on a given length $L$ from a
regular ``picket fence" spectrum of a harmonic oscillator.
${\mathcal N}(E)$ is the cumulative level number (the number
of levels with energy less than or equal to $E$). The angle
brackets in Eq. (\ref{eq:d3}) imply averaging over all values
of $i$, the location of the window of $L$ levels within the
spectrum. $A$ and $B$ are chosen so as to minimize $
\Delta^i_3(L)$; they are recalculated for each value of $i$,
the starting point of the fragment sliding along the
spectrum.

A series of evenly spaced levels would make ${\mathcal N}(E)$ a
regular staircase, then $\Delta_3(L)=1/12$. At the other extreme, a
classically regular system will lead to a quantum mechanical
spectrum with no level repulsion, the fluctuations will be far
greater, and $\Delta_3(L)=L/15$. One can also introduce another
useful statistic, the level number variance, $\Sigma^2(L)$,
\cite{guhr}. It is the variance in the number of levels found in an
interval of length $L$. After unfolding the spectrum, one expects
there to be $L \pm \sqrt{\Sigma^2(L)}$ levels in the interval.  For
a regular spectrum one has $\Sigma^2(L)=L$, while for a harmonic
oscillator spectrum it is zero. The relationship between
$\Sigma^2(L)$ and $\Delta_3(L)$ is given in \cite{pandey79} as
\begin{equation}
\Delta_3(L)=\frac{2}{L^4}\int^{L}_{0}(L^3-2L^2r+r^3)\Sigma^2(r)dr
.
\end{equation}

The asymptotic RMT result for the Gaussian Orthogonal
Ensemble (GOE) is
\begin{equation}
\Delta_3(L)=\frac{1}{\pi^2}\,\left[\log(2\pi
L)+\gamma-\frac{5}{4}-\frac{\pi^2}{8}\right]
\label{eq:d3th},
\end{equation}
with $\gamma$ being Euler's constant. We stress that this is
the RMT value for the {\sl ensemble average} of
$\Delta^i_3(L)$, not a spectral average. Putting in values we
get $\Delta_3(L)=(\log L-0.0678)/\pi^2$. For the GOE this
statistic increases very slowly with $L$, the levels are
crystalized into a rigid structure, hence the alternative
name ``spectral rigidity" for this statistic.

In this paper we are concerned with detailed properties of
the $\Delta_3(L)$ statistic and its use for the RMT analysis
of neutron resonance data. The main issues we will address
are the fraction, $x$, of missing levels, and the
uncertainties on the $\Delta_3(L)$ calculation. $\Delta_3(L)$
has been applied in this context before. In \cite{georgop81}
Monte Carlo calculations were used to see the effect of
missing levels on pure and mixed GOE spectra. They give
empirical graphs that can be used to get $x$, given
$\Delta_3(L)$ of a specific experimental spectrum. They also
give an empirical expression for the uncertainties in
$\Delta_3(L)$ that include effects of both sample size and
$x$. In \cite{brody} an expression for the uncertainties is
suggested, and we verify it here numerically. In
\cite{shriner92} the effect of sample size on $\Delta_3(L)$
was examined, and the level spacing distribution was deemed a
more useful statistic. We reexamine the question here of how
to compare $\Delta_3(L)$ calculated from a set of neutron
resonance data with RMT. We will give an exact method of
calculating $\Delta_3(L)$ for an unfolded spectrum, and a
consistent approach to comparing the experimental result with
the theoretical model.

\begin{figure}
\includegraphics[width=6.5in]{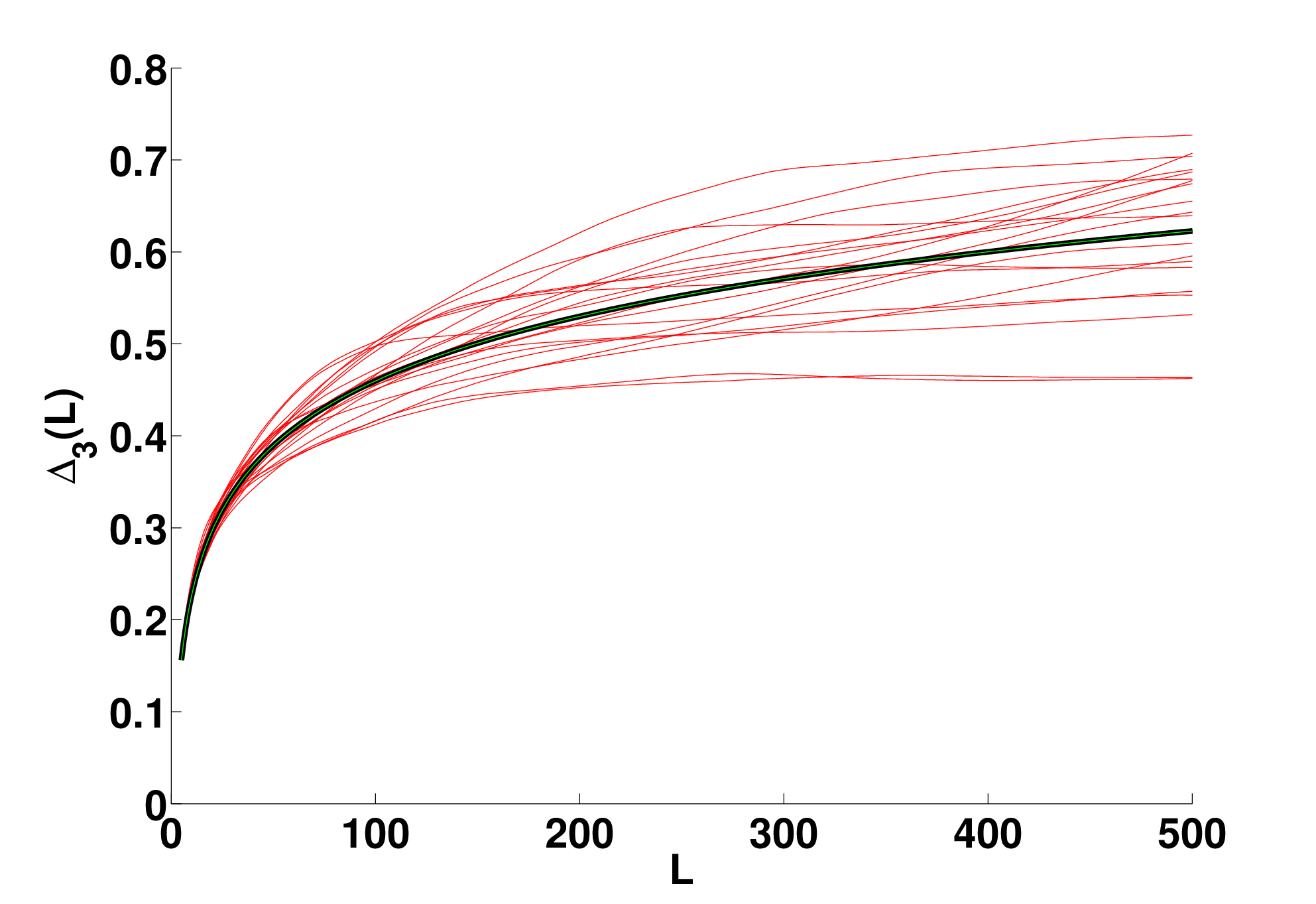}
\caption{\label{fig:manyspec50} (Color online) Calculated
$\Delta_3(L)$-statistic for an ensemble of 50 GOE spectra
with $N=4000$, results for 20 spectra are shown as thin lines
(red). The ensemble average, $\overline{\Delta_3(L)}$, for
all the 50 spectra lies on the theoretical thick (black)
curve, differing from it by $\approx 2\times 10^{-3}$ over
the whole range of $L$.}
\end{figure}

To illustrate the problem, see  Fig.~\ref{fig:manyspec50}, where
$\Delta_3(L)$ is shown for  20 out of  a set of 50 GOE spectra of
random GOE matrices of dimension $N=4000$. Notice the lines tend to
be quite smooth, but there is a considerable spread. The mean value
of $\Delta_3(L)$ for these 50 spectra is shown in green in
Fig.~\ref{fig:manyspec50}, but is not visible as it lies on the
theoretical curve, Eq. (\ref{eq:d3th}). This average of the red
lines is $\overline{\Delta_3(L)}$, the {\sl ensemble average} of
$\Delta_3(L)$. In this way we have recovered the theoretical result,
the difference between theory and calculation $\approx2\times
10^{-3}$ for the large range of $L$ shown. The source of the
discrepancy for a specific spectrum is simply the natural spread in
$\Delta_3(L)$ values for different spectra. This begs the question:
if a $\Delta_3(L)$ calculation on some experimental data gave one of
the lines in Fig.~\ref{fig:manyspec50} what conclusions could be
drawn about the purity of the spectra, missing levels, etc. We need
an ``uncertainty" to define a confidence interval centered on the
theoretical line, so we can make meaningful comparisons with the
data.

\section{\label{sec:d3disc} Ergodicity of  $\Delta_3(L)$}

The validity of a comparison between the spectral average of a
quantity with the theoretical ensemble average depends on the
quantity being {\sl ergodic}. For a clear and more detailed account
of this topic see \cite{brody}.

Consider an observable $X(E)$, which is some function of energy. In
RMT, this observable would be calculated by evaluating $X(E)$ for
fixed $E$ and averaging over an ensemble of spectra to get the
ensemble average, $ \overline{X(E)}$. The variance of $X(E)$ is
written $\text{Var}_e(X)=\overline{X(E)^2}-\overline{X(E)}\:^2$. We
will write the standard deviation as $\sigma_eX$, or simply
$\sigma_e$, the subscript indicating ensemble averaging. Dyson
\cite{dyson} derived the variance of $\Delta_3$ in GOE to be
$\text{Var}_e(\Delta) =1.169/\pi^4=0.110^2$, so $\sigma_e=0.110$. On
the other hand, experimentally one is dealing with an interval of
the spectrum over an energy range (determined by the experiment)
$[E, E+\Delta E ]$, and one calculates the spectral average within
that range, as a running average over the energy,
\begin{equation}
\langle X(E)\rangle=\frac{1}{\Delta E}\int^{E+\Delta
E}_{E}X(E') dE'.
\end{equation}
Ergodicity is equivalent to the statement $\langle
X(E)\rangle=\overline{X(E)}$.

\begin{figure}
\includegraphics[width=6.5in]{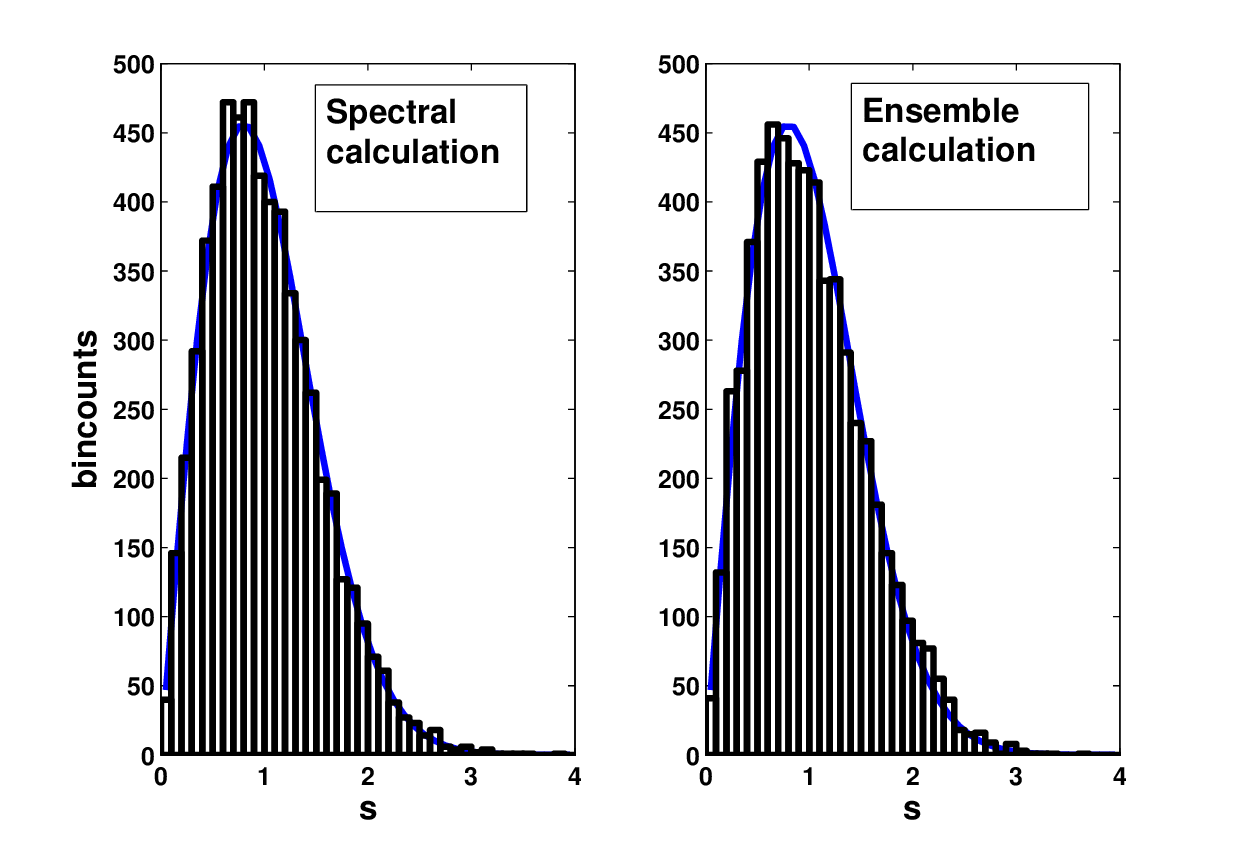}
\caption{\label{fig:pofsergodic}(Color online) The
distribution of level spacings, $s$, taken from one $N=6000$
GOE spectrum, left, and the distribution of the center level
spacing, $E_{300}-E_{299}$, taken from an ensemble of 6000
GOE spectra with $N=600$. The distributions are the same. The
statistic $s$ is ergodic.}
\end{figure}

Take, for example, the nearest neighbor level spacing $s =
E_{i+1}-E_i$. It is easy to verify that $s$ is ergodic, see
Fig.~\ref{fig:pofsergodic}. In this case the probability
density for $s$ is the same within a spectrum as it is in the
ensemble. The formal requirement for $X(E)$ to be ergodic is
that $\text{Var}_e \langle X(E)\rangle=\overline{\langle
X(E)\rangle^2}-\overline{\langle
X(E)\rangle}\:^2\rightarrow0$ as $\Delta E
\rightarrow\infty$. As it stands it is not particularly
useful, what we need is the behavior of $\text{Var}_e \langle
X(E)\rangle$ for finite data sets. Specifically, {\it we need
an expression for the uncertainty in the quantity after
replacing ensemble averaging with spectral averaging}. We
will refer to this quantity as $\sigma$, with $\text{Var}_e
\langle X(E)\rangle = \sigma^2$.

In application to $\Delta^i_3(L)$, the energy dependence is in $i$,
which indicates the location $E_i$ of the window of $L$ levels. The
spread in the individual lines in Fig.~\ref{fig:manyspec50} is
$\sigma$, while their average is
$\overline{\langle\Delta^i_3(L)\rangle}$, or
$\overline{\Delta_3(L)}$, which is the notation we will use. It is
$\sigma$ that will determine the sensitivity of $\Delta_3$ as a tool
for detecting missing levels. In a calculation of $\Delta_3(L)$ on a
spectrum of $N$ levels, we take a spectral average of
$\Delta_3^i(L)$, where the average is taken over the $N-L$ possible
locations $E_i$ of the window of $L$ levels. Brody {\sl et al.}
\cite{brody} discuss the situation where a quantity $X$ is
calculated over $p$ non-overlapping intervals. They suggest that
$\text{Var}_e\langle X \rangle_p = \text{Var}_e(X)/p$. They call
this the Poisson estimate. In the case of the $\Delta_3$ statistic,
there are $N/L$ non-overlapping intervals available for each $L$,
and the Poisson estimate would prescribe an uncertainty of $\sigma =
0.11 \sqrt{L/N}$. We will verify this numerically for the GOE. In
practice the intervals used in the calculation overlap, as we allow
$i$ to take all available $N-L$ values. The values of
$\Delta_3^i(L)$ are highly correlated in this case however, and the
Poisson estimate is still good. It is clear from this result that
$\Delta_3$ is a more sensitive statistic for small values of $L/N$.

In the case of $m$ independent spectra, superimposed in proportions
$f_1,\, f_2,\,\dots f_m$, and letting $\Delta_{3m}(L)$ be the
spectral rigidity of the $m^{\textrm{th}}$ sub-spectrum, we have
\cite{dyson} $\Delta_3(L)=\sum_{i=1}^m \Delta_{3m}(f_i L)$. If the
$m$ spectra are all from the GOE, the ensemble variance is
$\sigma_e^2 = (0.110\,m)^2$, and the Poisson estimate  then gives
$\sigma = 0.110\,m\,\sqrt{L/N}$. Based on this number $\Delta_3(L)$
can be used to distinguish between spectra with $m=1$ and $m=2$
independent sequences present. However this statistic is not
sensitive to the actual mixing fractions. We have assumed here that
the proportions $f_1,\, f_2,\,\dots f_m$ are independent of energy.
This may not be the case, in neutron resonances, the fraction of
intruder p-wave resonances may grow with energy.

\begin{figure}
\includegraphics[width=4in]{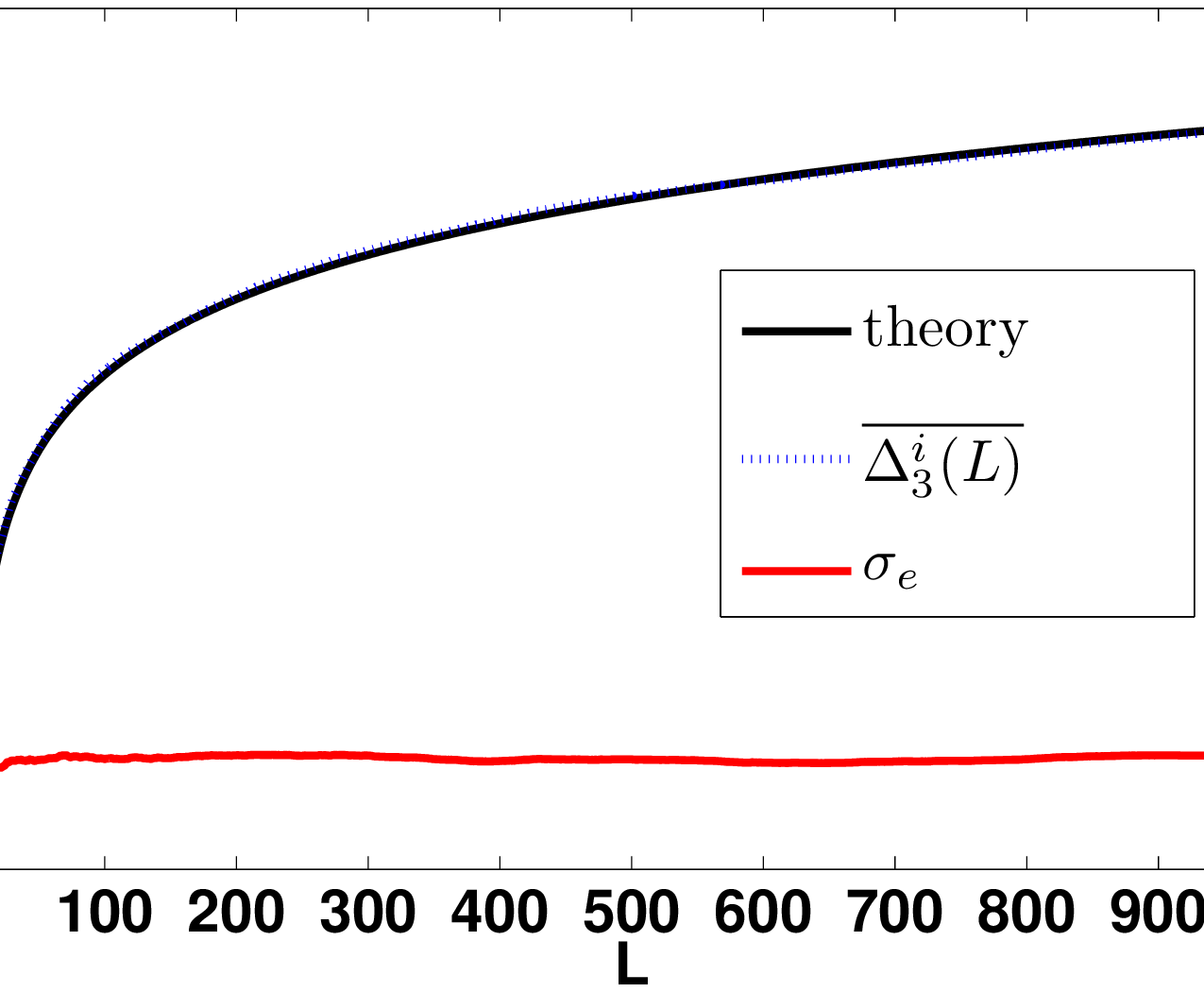}
\caption{\label{fig:ergodic} (Color online) The dotted (blue)
curve is $\overline{\Delta^i_3(L)}$, the ensemble average of
$\Delta^i_3(L)$, for $i=1200-\frac{L}{2}$; the ensemble of
1000 spectra with $N=2400$ was used. The ensemble average is
indistinguishable from the RMT prediction, solid  (black)
line. The lower (red) curve is $\sigma_e$, the standard
deviation of $\overline{\Delta^i_3(L)}$. It agrees with the
RMT result of 0.11.  }
\end{figure}

To calculate the ensemble average,
$\overline{\Delta^i_3(L)}$, 1000 matrices of dimension 3000
were made. Each of these spectra was unfolded. The value of
$i=\frac{N}{2}-\frac{L}{2}$ was chosen to locate the window
of $L$ levels squarely in the middle of the spectrum,
$\Delta_3^i(L)$ was calculated, and the resulting
$\overline{\Delta^i_3(L)}$ matched the RMT result, see
Fig.~\ref{fig:ergodic}. As a double check, the process was
repeated for the value of $i=\frac{N}{4}-\frac{L}{2}$, which
locates the window of $L$ levels on the left edge of the
spectrum. A discrepancy here would undermine our unfolding
procedure, see Sec. \ref{sec:d3}, but the results were
identical.

\section{\label{sec:d3}Calculation of  $\Delta_3(L)$}

To realize the GOE, we generated random matrices with
normally distributed matrix elements, and
\begin{equation}
P(H_{i\neq j})=\frac{1}{\sqrt{2 \pi
\sigma^2}}\,e^{-\frac{H_{ij}^2}{2 \sigma^2}}, \quad
P(H_{ii})=\frac{1}{\sqrt{4 \pi
\sigma^2}}\,e^{-\frac{H_{ii}^2}{4 \sigma^2}}
\end{equation}
for the off-diagonal and diagonal elements, respectively. We chose
$\sigma=1$.  Each of the matrices has an approximately semicircular
level density, with $\rho(E)=\sqrt{4N-E^2},\;\text{for }|E|\leq
2\sqrt{N},\;0 \text{ otherwise}$ (actually there are deviations from
the semicircle at the edges, see Mehta \cite{mehta}).

In order to apply the results of RMT to real data, the
empirical spectrum must be unfolded \cite{guhr,brody}
separating the fluctuations of the spectra from secular
behavior and expressing all energies in units of the local
average level spacing. This is achieved by first extracting
the cumulative level density ${\mathcal N}(E)$, which will be
a staircase function, from the raw data, and fitting it to a
smooth function, $\xi(E)$, either numerically, or
analytically. Then, using this function, the $j^{th}$ level
of the unfolded spectrum is simply $\xi(E_j)$. The resulting
unfolded spectrum has a uniform level density $\rho(E)=1$ so
that data from high density regions can be compared with data
from low density regions. Furthermore, spectra from very
different physical systems can be compared.

Integrating the semicircular GOE level density gives
\begin{eqnarray}
\xi(E)&=  \int_{-\infty}^{E}\rho
(E') dE'\nonumber\\
&=\frac{1}{\pi}
N~\tan^{-1}\left(\frac{E}{\sqrt{4N-E^2}}\right)+\frac{1}{4\pi}\,
E\sqrt{4N-E^2}+\frac{1}{2}N.\label{eq:NofE}
\end{eqnarray}
To decide between using the analytical form, or a numerical
fit, the curve fitting tool in MATLAB was used to get the
best values for $a$, $b$, and $c$ in the parametrization
\begin{equation}
\xi(E)= a~\frac{1}{\pi}
N\,\tan^{-1}\left(\frac{E}{\sqrt{4N-E^2}}\right)
+b~\frac{1}{4\pi}E\sqrt{4N-E^2}+c~\frac{1}{2}N.\label{eq:enfit}
\end{equation}
The result is, with 95\% confidence bounds,
$a=0.9992~(0.9992,0.9993)$, $b=1~(1,1.001)$,and $c=1~(1,1)$.
Using these values for a test spectrum with $N=500$ made a
difference in $\Delta_3(L)$ of $10^{-3}$ at $L=250$, so the
theoretical (semicircle) result, Eq. (\ref{eq:NofE}), was
used in all the calculations of $\Delta_3(L)$ for GOE
spectra, without any fitting. To realize ensembles of GOE
spectra of various spectrum size, $N$, we first made an
ensemble of 3000 unfolded GOE spectra with dimension 3000.
Getting a spectrum of  size $N$ is now a matter of taking the
middle $N$ eigenvalues from one of these. In what follows,
all calculations are performed on unfolded spectra.

We rewrite for convenience the definition of $\Delta_3(L)$
for a spectrum:
\begin{equation}
\Delta_{3}(L) = \left\langle {\rm min}_{A,B}\;
\frac{1}{L}\;\int^{E_i+L}_{E_i}dE'\,[\;{\mathcal
N}(E')-AE'-B]^{2}\; \;\right\rangle\:.
\label{eq:d3sect2}
\end{equation}
The  integration is over a window of the spectrum, of length
$L$ levels, starting at energy $E_i$, the $i^{th}$ level.
Note that some authors choose the limits of integration to be
$E_{i-\frac{L}{2}}, E_{i+\frac{L}{2}}$.  The integral is
evaluated for every starting energy $E_i$, with $1\leq i \leq
N-L$. $A$ and $B$ are chosen so as to minimize the integral
for each position of the window of $L$ levels, i.e. for each
value of $i$. The precise values of $A$ and $B$ will be given
in terms of $E_i$, no numerical minimization is necessary.

Substituting ${\mathcal N}(E)=i, E_i \leq E < E_{i+1}$, into
(\ref{eq:d3sect2}), our job is reduced to finding the mean of the
quantity $\Delta_3^i(L)$. In evaluating
\begin{equation}
\Delta_3^i(L)=\frac{1}{L}\;\int^{E_i+L}_{E_i}dE'\,[\;{\mathcal
N}(E')-AE'-B]^{2}\; ,\nonumber
\end{equation}
consider the integral between two adjacent levels,
\begin{equation}
\Delta_3^i(L)=
\frac{1}{L}\;\sum^{i+L-1}_{j=i}\int^{E_j+1}_{E_j}dE'\,(j-AE'-B)^{2}
= \frac{1}{L}\times(C+VA^2+WA+XAB+YB+ZB^2), \label{eq:d3iexplicit}
\end{equation}
where

\begin{eqnarray}
 \nonumber C &=& \sum^{i+L-1}_{j=i}j^2(E_{j+1}-E_j), \\
 \nonumber V &=& \frac{1}{3}(E_{i+L}^3-E_i^3), \\
 \nonumber W &=& \sum^{i+L-1}_{j=i}-j(E_{j+1}^2-E_j^2), \\
 \nonumber X &=& (E_{i+L}^2-E_i^2),\nonumber \\
 \nonumber Y &=& \sum^{i+L-1}_{j=i}-2j(E_{j+1}-E_j), \\
 \nonumber Z &=& (E_{i+L}-E_i).
\end{eqnarray}

Using the constraints $\partial (\Delta_3^i)/\partial A = 0$ and
$\partial (\Delta_3^i)/\partial B = 0$, we come to the following
expressions for $A$ and $B$ that minimize $\Delta_3^i(L)$:
\begin{equation}
A=\frac{XY-2WZ}{4VZ-X^2}, \quad B= \frac{WX-2VY} {4VZ-X^2}.
\end{equation}
Given an unfolded spectrum, $\Delta_3(L)$ can be calculated exactly.
Bohigas and Giannoni derived a similar expression for the case of
$L=N-1$ in \cite{bohigas75}. It is interesting to note that many
early investigations of $\Delta_3(L)$ deal with just this special
case, and hence distinction between a spectral average and an
ensemble average didn't arise, there being only one value of the
statistic per spectrum.

\section{\label{sec:sig}Calculation of $\sigma(L)$}

\begin{figure}
\includegraphics[width=4in]{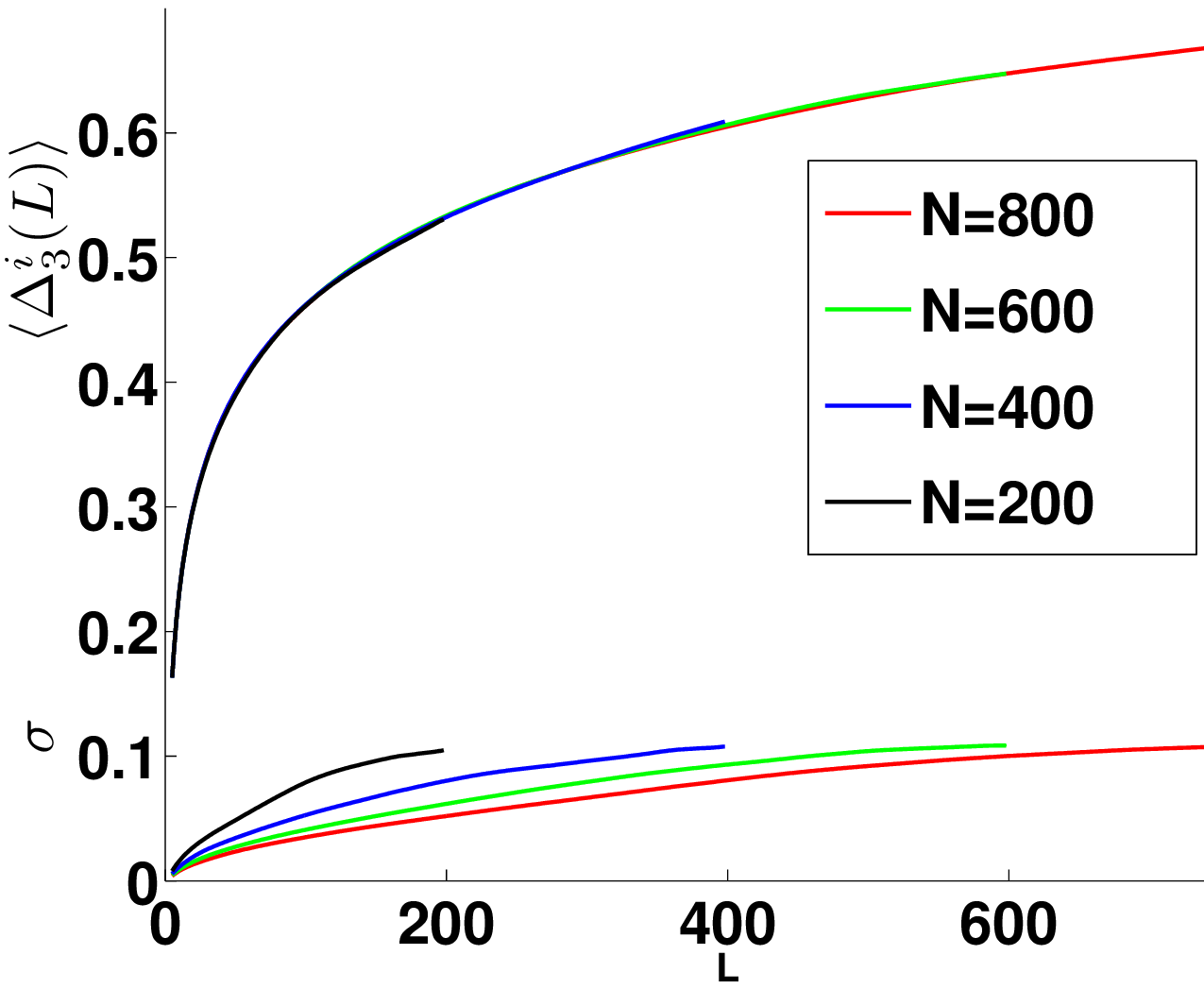}
\caption{\label{fig:sigma}(Color online)  The ensemble average,
$\overline{\Delta_3(L)}$ vs. L is plotted for various values of $N$.
Each ensemble had 1000 matrices. The lower lines are the standard
deviations of the corresponding 1000 values of $\Delta_3(L)$. Notice
that for the maximum value of $L$, $\sigma=0.11$, which is the
ensemble result. This is expected, as there is only one such value
per spectrum.}
\end{figure}

To calculate $\sigma$, the uncertainty for $\Delta_3(L)$ of a
specific GOE spectrum, we take 1000 GOE spectra, each with $N$
levels, and calculate $\Delta_3(L)$ for each. $\sigma$ is the
standard deviation of those 1000 numbers for each $L$, and this is
what we should use as the uncertainty for $\Delta_3(L)$ for that
value of $L$. In Fig.~\ref{fig:sigma} we show the
$\overline{\Delta_3(L)}$ and $\sigma$ for a range of $N$. The most
important feature is that $\sigma < \sigma_e$ for small values of
$L$. When $L=N$ then we have one number per spectrum, and $\sigma =
\sigma_e$ as expected.

\begin{figure}
\includegraphics[width=4in]{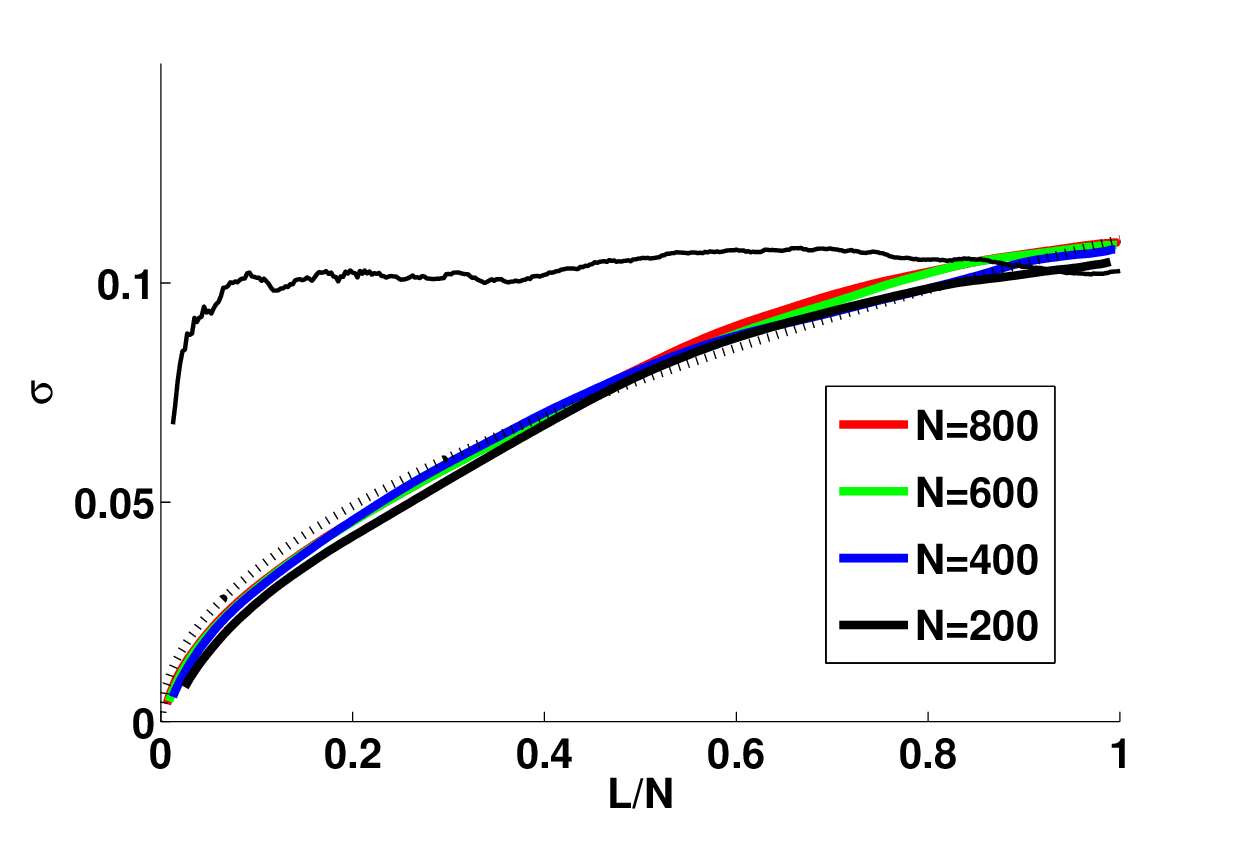}
\caption{\label{fig:sigmaLN} (Color online) Here we plot $\sigma$
from Fig.~\ref{fig:sigma} vs. $L/N$, for various ensembles. The top
(black) line is $\sigma_e$, in agreement with the RMT result of a
constant 0.11 . The dashed line is $\sigma_e\sqrt{L/N}$.}
\end{figure}

To compare different size spectra, we plot $\sigma$ vs. $L/N$, in
Fig.~\ref{fig:sigmaLN}.  A numerical fit to the range $0\le L/N \le
0.2$ gives
\begin{equation}
\sigma = 0.1126 (L/N)^{0.5003} - 0.0045 \approx \sigma_e
\sqrt{L/N}.\nonumber
\end{equation}
This is a verification of the Poisson estimate. The dashed
line in Fig.~\ref{fig:sigmaLN} is $\sigma = \sigma_e
\sqrt{L/N}$. The Poisson estimate was stated for the case of
non-overlapping intervals. In Eq. (10) this would mean
$i=1,\,L+1,\,2L+1,\dots$. In our calculations all values of
$i$ were used, this is equivalent to doing $L$ sets of
calculations for non-overlapping intervals, with the initial
values of $i$ in each set going from 1 to $L$, and then
taking the average of these $L$ numbers, so the Poisson
estimate is still valid. Note that these ``uncertainties" are
not to be confused with any experimental uncertainty or
computational issue. Given an unfolded data set, the
$\Delta_3(L)$ statistic is calculated exactly, as was shown
in Sec. \ref{sec:d3}.

\begin{figure}
\includegraphics[width=4in]{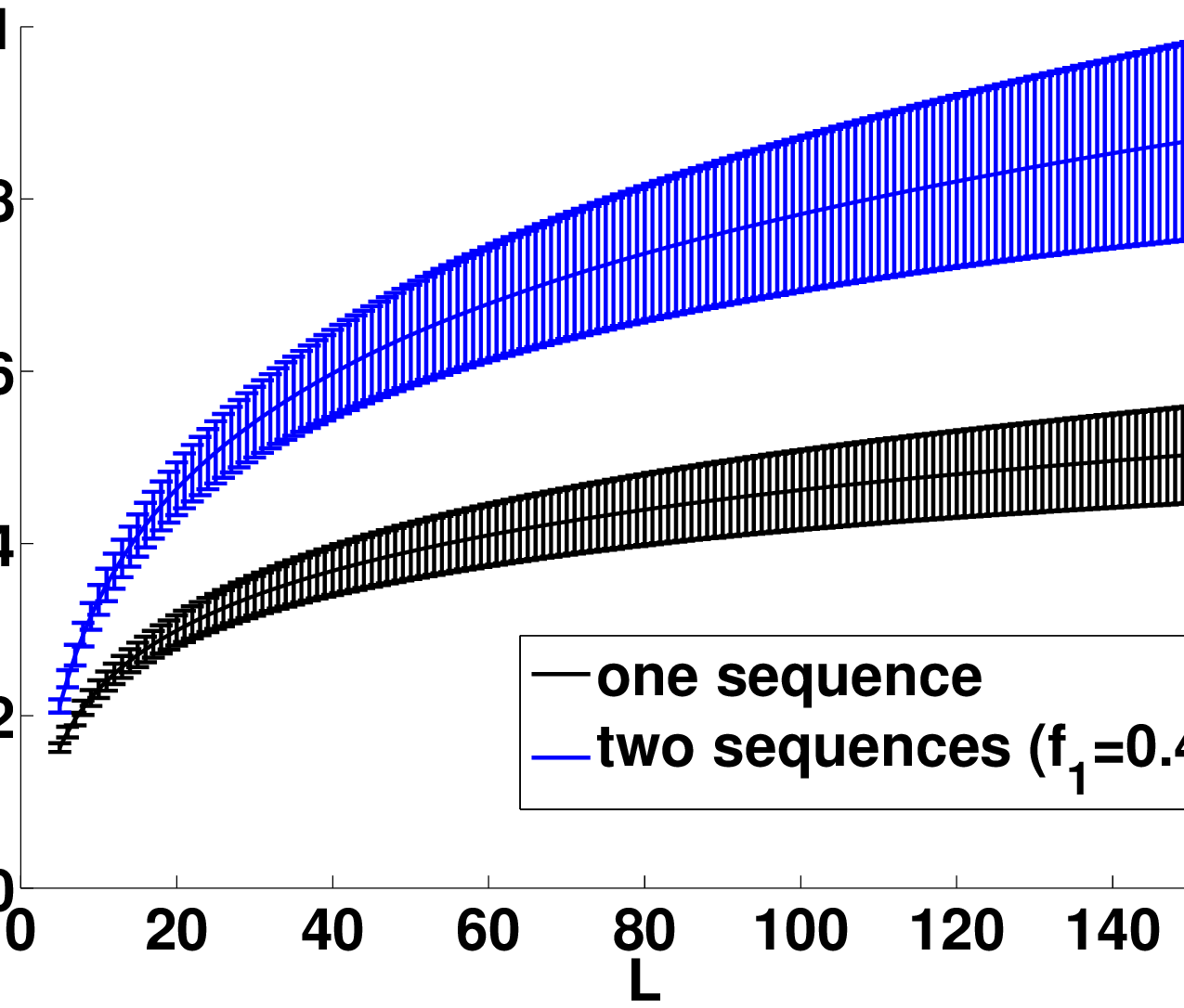}
\caption{\label{fig:empire} (Color online)
$\overline{\Delta_3(L)}$ is plotted with uncertainties for
the cases of an ensemble of 500 pure GOE spectra,  lower
lines (black), and a 500 mixed GOE spectra, upper lines
(blue). In the mixed case, each spectrum was a superposition
of 2 pure GOE spectra in proportions $f_1=0.4\,,f_2=0.6$. In
both ensembles, the spectra had $N=500$. The uncertainties
are empirical, being the standard deviation of the 500 values
of $\Delta_3(L)$ at each value of $L$.}
\end{figure}

For the case of a mixture of $m=2$ independent GOE spectra, the
ensemble result is $\sigma_{e}= 0.22$. A calculation of $\sigma$ for
500 spectra each with 500 levels, made by mixing two GOE spectra,
with $f_1=0.4$, gives $\sigma = 0.239 (L/N)^{0.522} - 0.014$, which
is consistent with $\sigma = 2~\sigma_e \sqrt{L/N}$. In
Fig.~\ref{fig:empire} we show $\Delta_3(L)$ with the empirical
uncertainties. The statistic can clearly distinguish between $m=1$
and $m=2$ spectra, and it is most sensitive for small values of $L$.
$\Delta_3(L)$ is not sensitive to the actual value of $f_1$, and
cannot be used to distinguish between $f_1=0.25$ and $f_1=0.4375$.
These are the values for the fraction densities one would expect for
neutron resonances on target nuclei with spin $j_{{\rm tar}}=1/2$
and $7/2$, based on the $2j+1$ degeneracies of the resulting
compound nucleus spins $j_{{\rm tar}}\pm 1/2$. The statistic is
sensitive to $p$-wave neutrons in the beam being mislabeled as
$s$-wave for spin-1/2 targets, as this would introduce multiple
sequences of levels into the data set.

\begin{figure}
\includegraphics[width=3.5in]{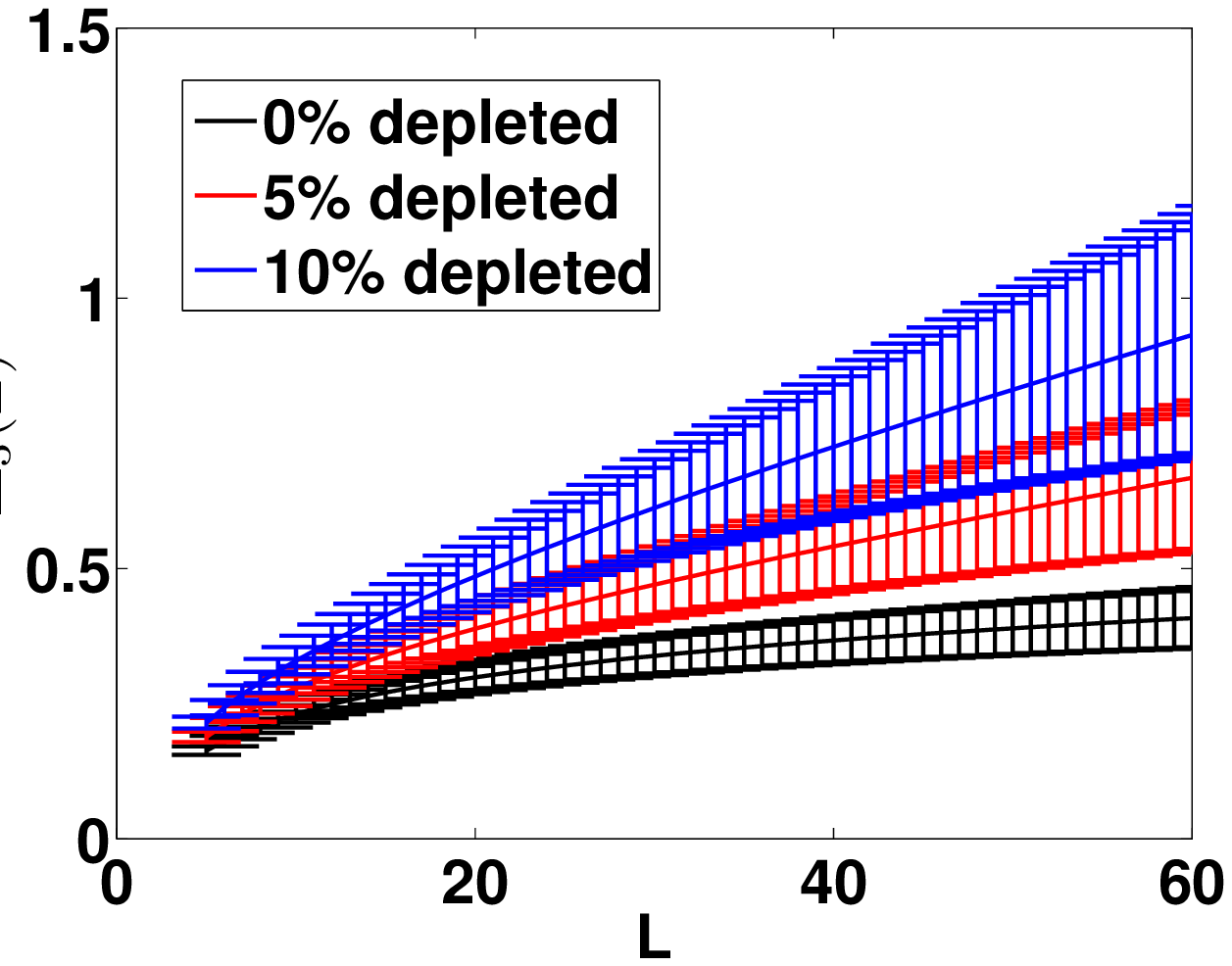}
\caption{\label{fig:sigmadep200} (Color online)
$\overline{\Delta_3(L)}$ vs. $L$ is plotted for depleted
spectra size, $N=200$, with uncertainties. Lower line has
$x=0\%$ depletion, middle line has $x=5\%$, and the upper
line has $x=10\%$.}
\end{figure}

\begin{figure}
\includegraphics[width=3.5in]{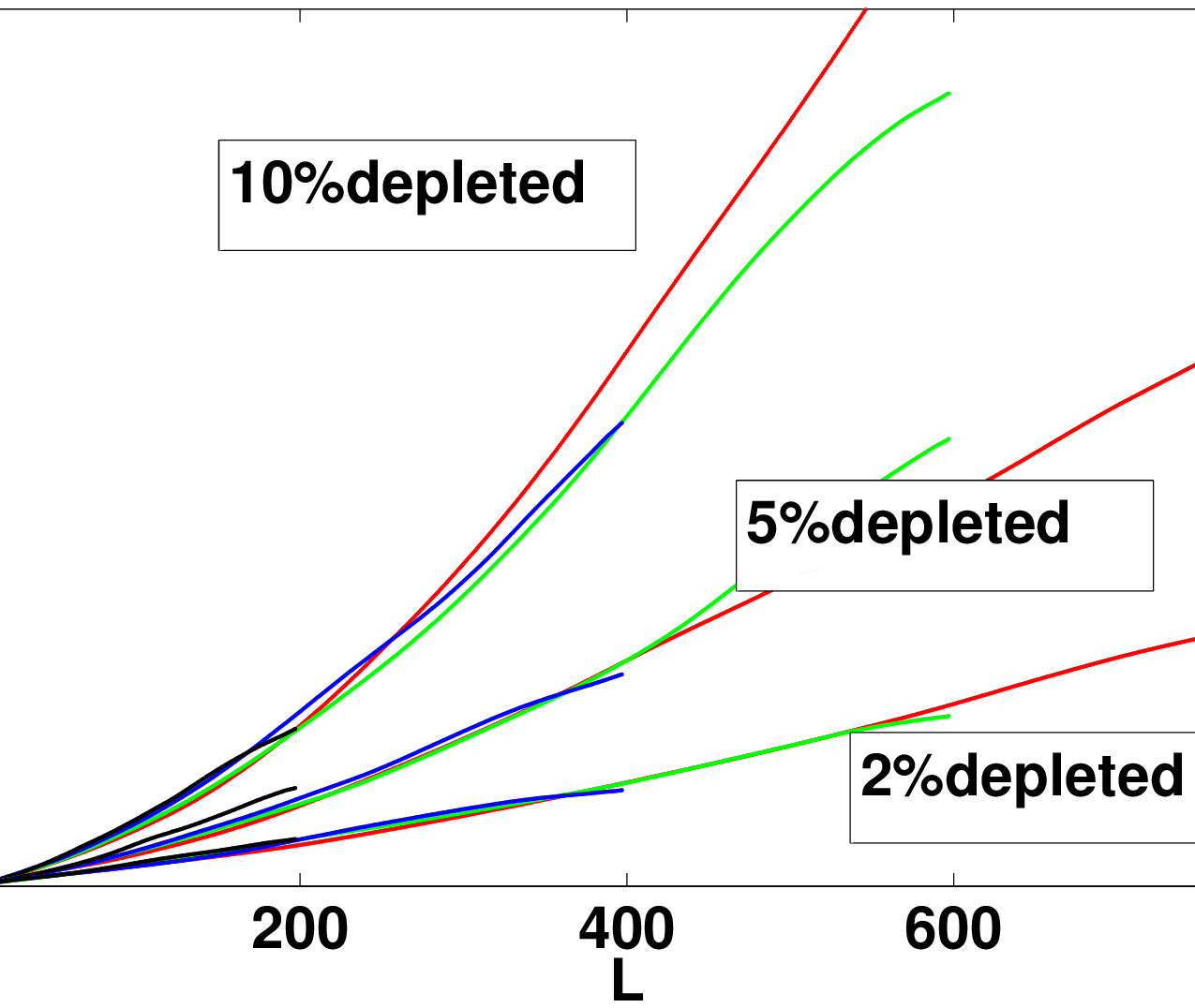}
\caption{\label{fig:sigma_sLNdep} (Color online) $\sqrt{N}\sigma$
vs. $L$ is plotted for depleted spectra of dimension $N$. The values
of $N$ are 200, 400, 600, and 800, $L$ goes out to its maximum value
of $N-1$.}
\end{figure}

To determine the completeness of a spectrum with
$\Delta_3(L)$, we need to know its behavior for GOE spectra
with a fraction of levels, $x$, depleted. We denote this by
$\Delta_3(L;x)$, and it was found empirically for ensembles
of 500 spectra with dimension $N$ and depletion $x$. The
dimension, $N$, refers here to the number of levels {\it
after} the fraction $x$ was randomly removed. The ensemble
average of $\Delta_3(L;x)$ and its standard deviation
$\sigma(N,L;x)$ were calculated. The results are shown in
Fig.~\ref{fig:sigmadep200} for $N=200$, with uncertainties.
We see from Fig.~\ref{fig:sigma_sLNdep} that the uncertainty
here has the form $\sigma_x(N,L;x)=f(L)/\sqrt{N}$. The
$1/\sqrt{N}$ dependence of $\sigma_x(N,L;x)$ means that the
statistic is a less sensitive measure of $x$ for lower $N$. A
fit to $f(L)$ was made for a range of $x$, to be used in
practical analysis of the neutron resonance data.

Now we have the tools necessary to compare a real spectrum with
depleted GOE spectra in a meaningful way. The best value for $x$
from the data will be the one that minimizes
\begin{equation}
\chi^2(x) = \sum^{L_{max}}_{L_{min}}
\frac{[\Delta_3(L)-\Delta_3(L;x)]^2}{\sigma(N,L;x)^2}.
\end{equation}

For practical purposes we need an estimate of the error in
$x$ from this method. To this end, we calculated the average
value of $x$, and its standard deviation, for 1500 GOE
spectra with $N=100$ and 300 levels. The sets were made by
randomly deleting 3 levels from a spectra of 103 levels, 5
out of 105, 9 out of 109, and 11 out of 111, to get spectra
with $x= 2.91\%$, 4.76\%, 8.26\%, and 9.9\% respectively. The
results are in Table \ref{tab:d3error}. Our method gave good
agreement for the value of $x$, but the uncertainties were of
the same order as $x$, for example, with $x=4.76\%$, we get a
value of $(4.89 \pm 2.81)\%$. However, when we tried the
method for $N=300$, the uncertainty dropped by a factor of
nearly $\sqrt{3}$ to 1.66\%, which is to be expected.

\begin{table}
\caption{\label{tab:d3error}The errors in $x$, the fraction
of missed levels, using the $\Delta_3$ statistic. The tests
were run for 1500 depleted spectra of size $N=100$ and 300.
The mean value $\overline{x}$, and the standard deviation,
$\sigma$ are given. }
\begin{ruledtabular}
\begin{tabular}{llllll}
$x$ & $\overline{x} \,(N=100)$ & $\sigma \,(N=100)$ & $\overline{x} \,(N=300)$ & $\sigma \,(N=300)$ \\
\hline
    2.91\% & 3.14\% & 2.50\% & 2.71\% & 1.57\%\\
    4.76\% & 4.89\% & 2.81\% & 5.00\% & 1.66\%\\
    8.26\% & 7.95\% & 3.13\% & 7.94\% & 1.74\%\\
    9.91\% & 9.90\% & 3.06\% & 10.0\% & 1.77\%\\
\end{tabular}
\end{ruledtabular}
\end{table}

\begin{figure}
\includegraphics[width=3.5in]{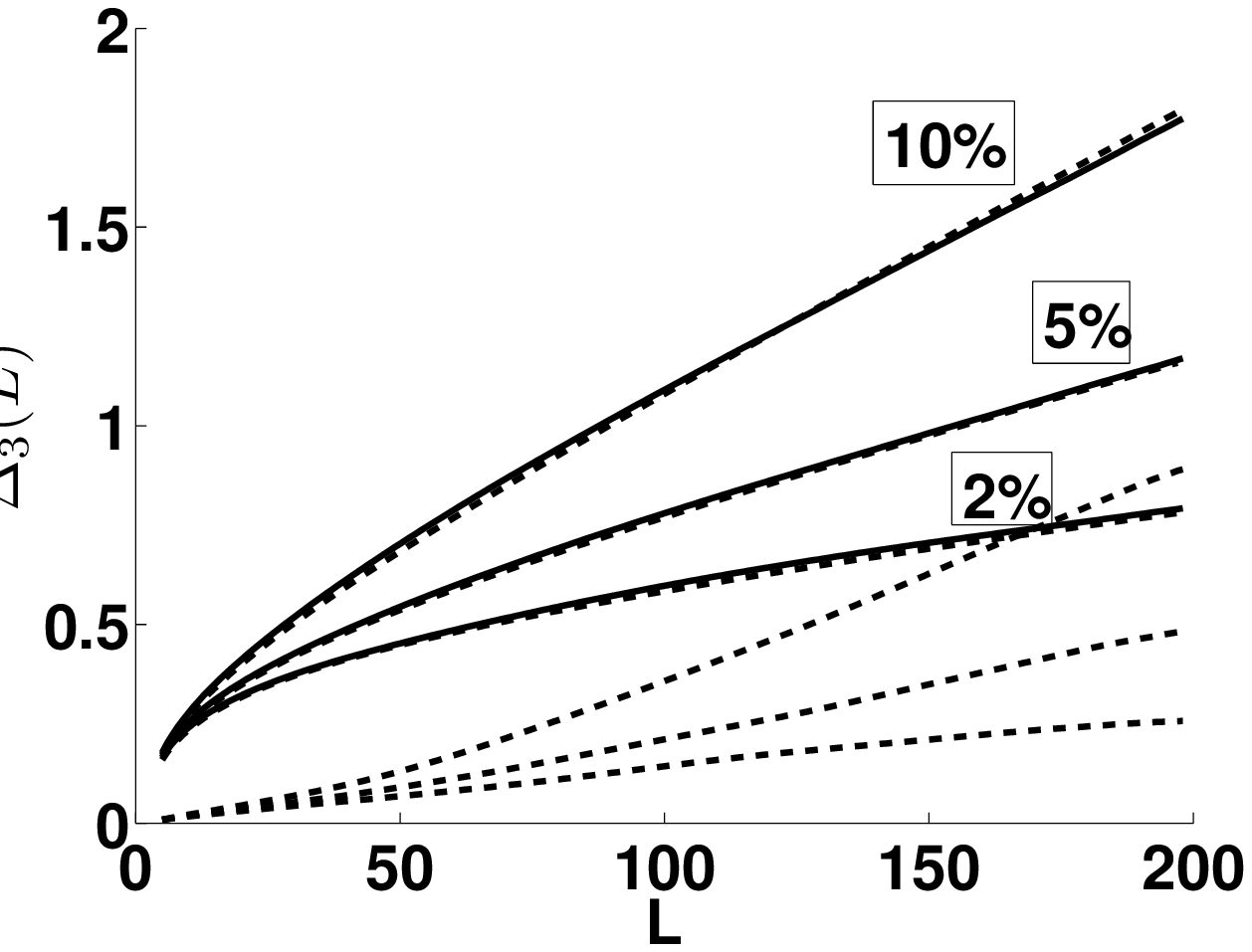}
\caption{\label{fig:intruder} Here $\overline{\Delta_3(L)}$
is shown for contaminated GOE spectra, upper solid lines. The
level of contamination, $x$, is labeled. Upper dashed lines
are the RMT predictions. The dashed lines at the bottom of
the plot are the standard deviations, $\sigma$, of
$\Delta_3(L)$ for $x=2\%,\, \text{the lowest of the three,
to}\,10\%,\, \text{the highest of the three}.$ }
\end{figure}

An experimental spectrum can be contaminated by intruder
levels. In the case of $s$-wave neutron resonance data, this
could be from the capture of $p$-wave neutrons. To examine
the effect of contamination on $\Delta_3(L)$, ensembles of
1000 unfolded GOE spectra with $N=200(1-x)$ were made. Each
spectrum was stretched by a factor of $1/(1-x)$ and then $200
x$ random numbers, uniformly distributed on the interval
$[0,200]$, were added. The resulting spectrum had a uniform
level density, $\rho(E)=1$, a size $N=200$, and a level of
contamination of $x\%$. In Fig.~\ref{fig:intruder} we see the
results for $x=2\%,\,5\%,\, \text{and}\; 10\%$ (solid lines).
The results agree well with the RMT prediction of
$\Delta_3(L)=(1-x)[(\log L-0.0678)/\pi^2]+x\,L/15$ (dashed
lines).

\begin{figure}
\includegraphics[width=4in]{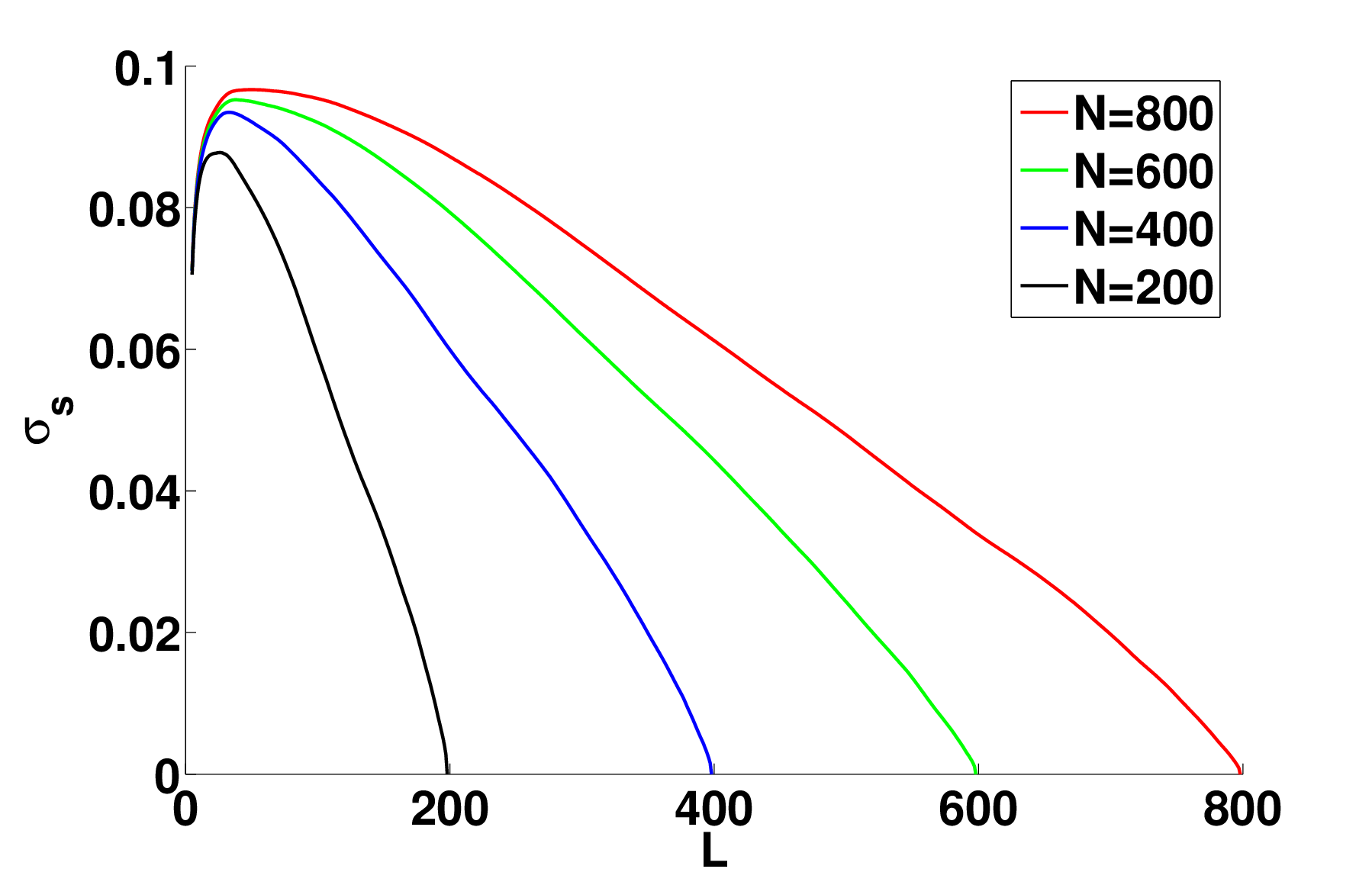}
\caption{\label{fig:sigma_s} (Color online) The standard
deviation of $\Delta_3^{i}(L)$ within a spectrum, $\sigma_s$,
vs. $L$ is plotted for various values of $N$. The linear
parts of the plot are well described by
$\sigma_s=\sigma_e(N-L)$}.
\end{figure}

\begin{figure}
\includegraphics[width=4in]{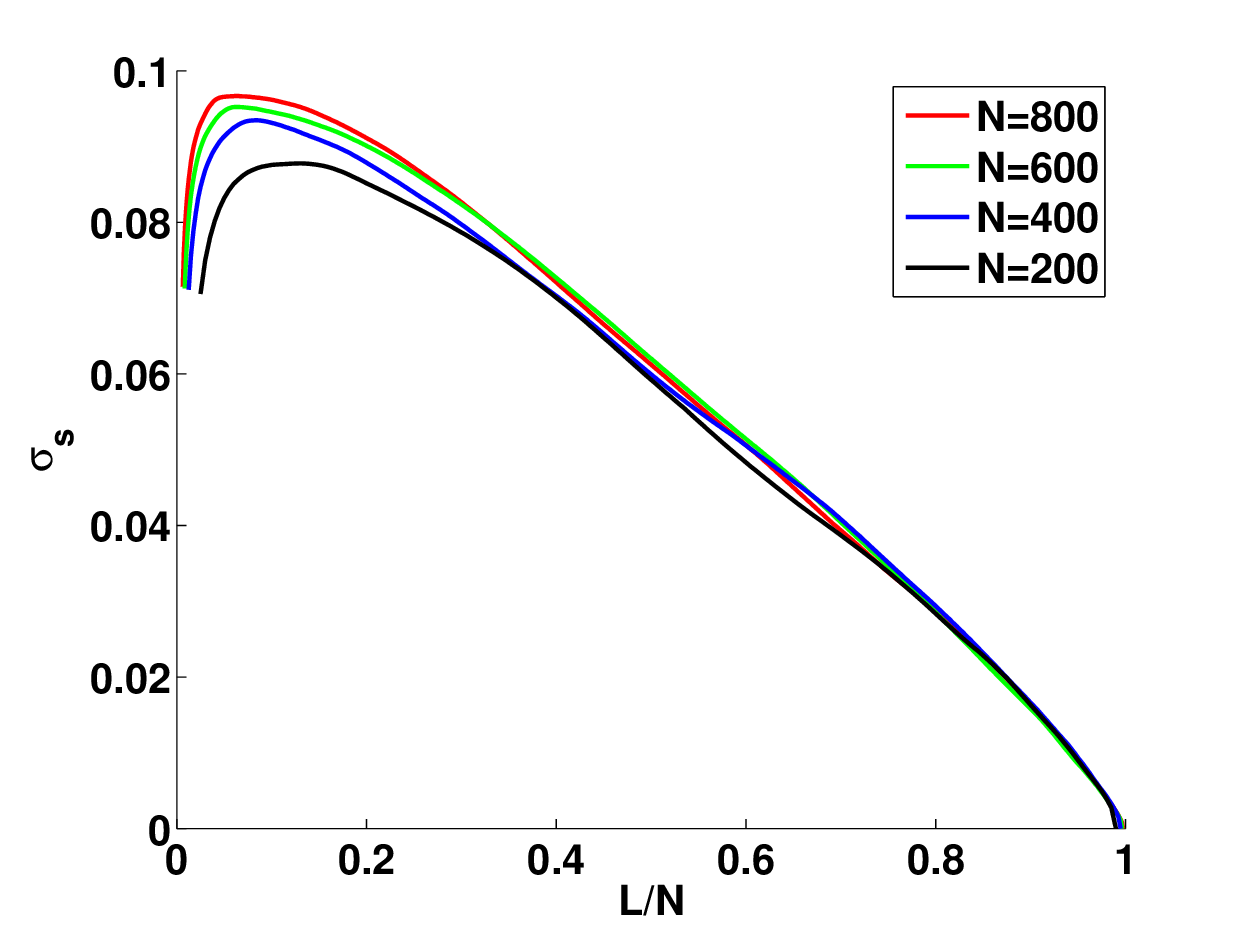}
\caption{\label{fig:sigma_sLN} (Color online) $\sigma_s$ vs.
$L/N$ is plotted for various values of $N$. The linear parts
of the plot are well described by
$\sigma_s=\sigma_e(1-L/N)$}.
\end{figure}

\begin{figure}
\includegraphics[width=4in]{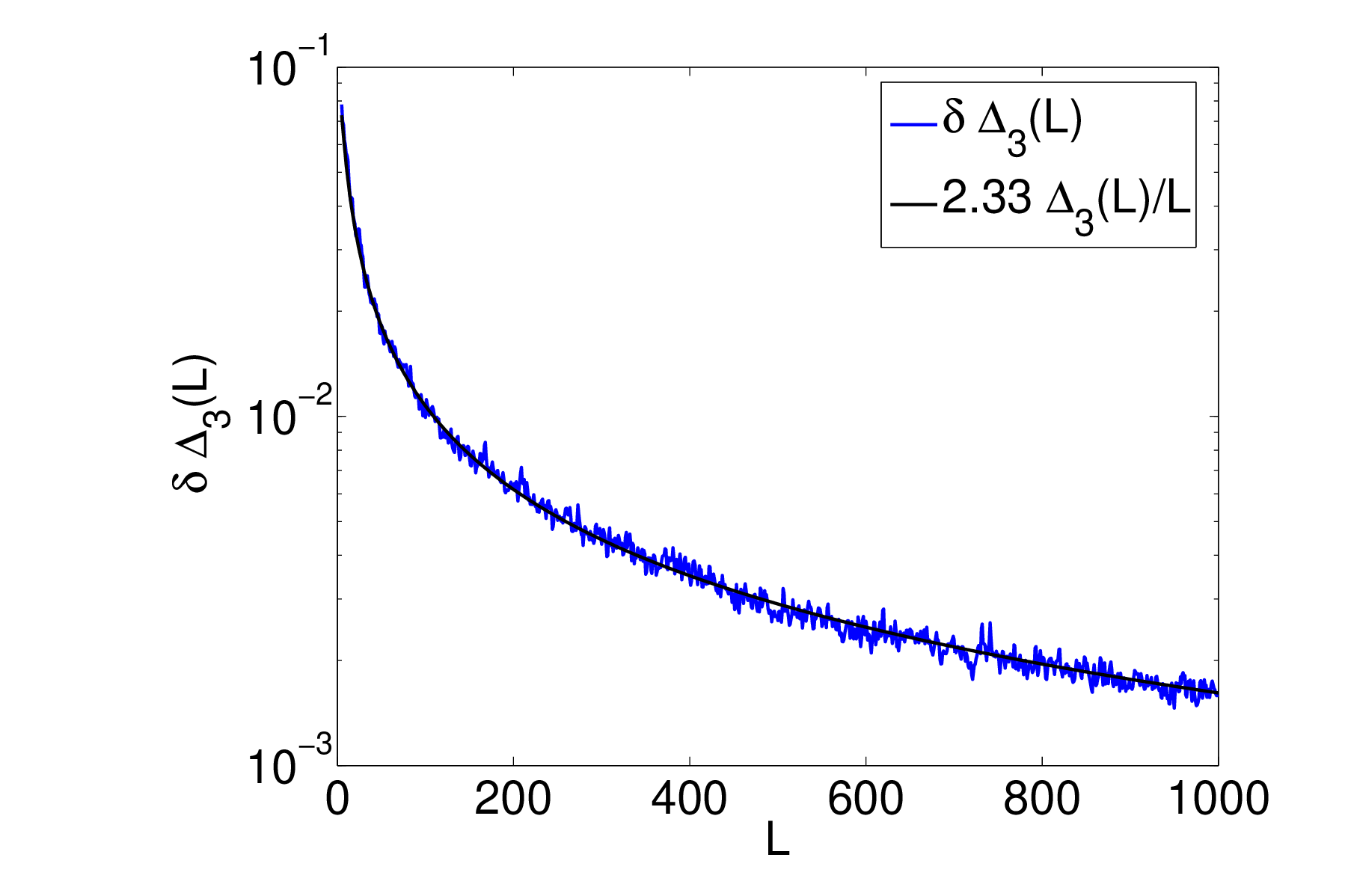}
\caption{\label{fig:deltaDelta} (Color online) The quantity
$\delta \Delta(L)$, as defined in the text, is plotted on a
log-linear graph. It is well approximated by $2.3303
\Delta_3(L)/L$ (smooth line).}
\end{figure}

We have discussed the spectral average of $\Delta^i_3(L)$, but what
about its spread within a spectrum? We call the standard deviation
of $\Delta^i_3(L)$ for a given spectrum $\sigma_s$. In
Fig.~\ref{fig:sigma_s} we see the average of $\sigma_s$ as a
function of $L$ for different spectra sizes, $N$. It is immediately
clear that $\sigma_s$ is less than $\sigma_e$. Within one spectrum,
we expect a smaller spread in the values of $\Delta^i_3(L)$ because
close values in $i$ mean the windows of $L$ levels overlap, the
corresponding values of $\Delta^i_3(L)$ would be correlated, and
$\sigma_s$ should get smaller as $L\rightarrow N$. A plot of
$\sigma_s$ versus $L/N$ strongly suggests the falloff is linear,
with $\sigma_s=\sigma_e(N-L)$, see Fig.~\ref{fig:sigma_sLN}. It is
not obvious that the correlations between overlapping windows of
levels would give this linear behavior. To gain insight into the
correlation between $\Delta_3^i(L)$ and $\Delta_3^{i+1}(L)$, we
examined the ensemble average of the square of the difference
between these two quantities, specifically, $\delta \Delta(L) =
\sqrt{\overline{[\Delta_3^{i+1}(L)-\Delta_3^{i}(L)]^{2}}}$, the
variation of $\Delta_3^{i}(L)$ with respect to $i$. It is expected
that this quantity should be closely related to $\Delta_3(L)$, and
it certainly decreases rapidly, $\delta \Delta(L) \propto
\Delta_3(L)/L$. The constant of proportionality is $\langle\frac{L
\; \delta \Delta(L)}{\Delta_3(L)}\rangle_L=2.3303$, where the
average was taken over all values of $L$, see
Fig.~\ref{fig:deltaDelta}.

\begin{figure}
\includegraphics[width=4in]{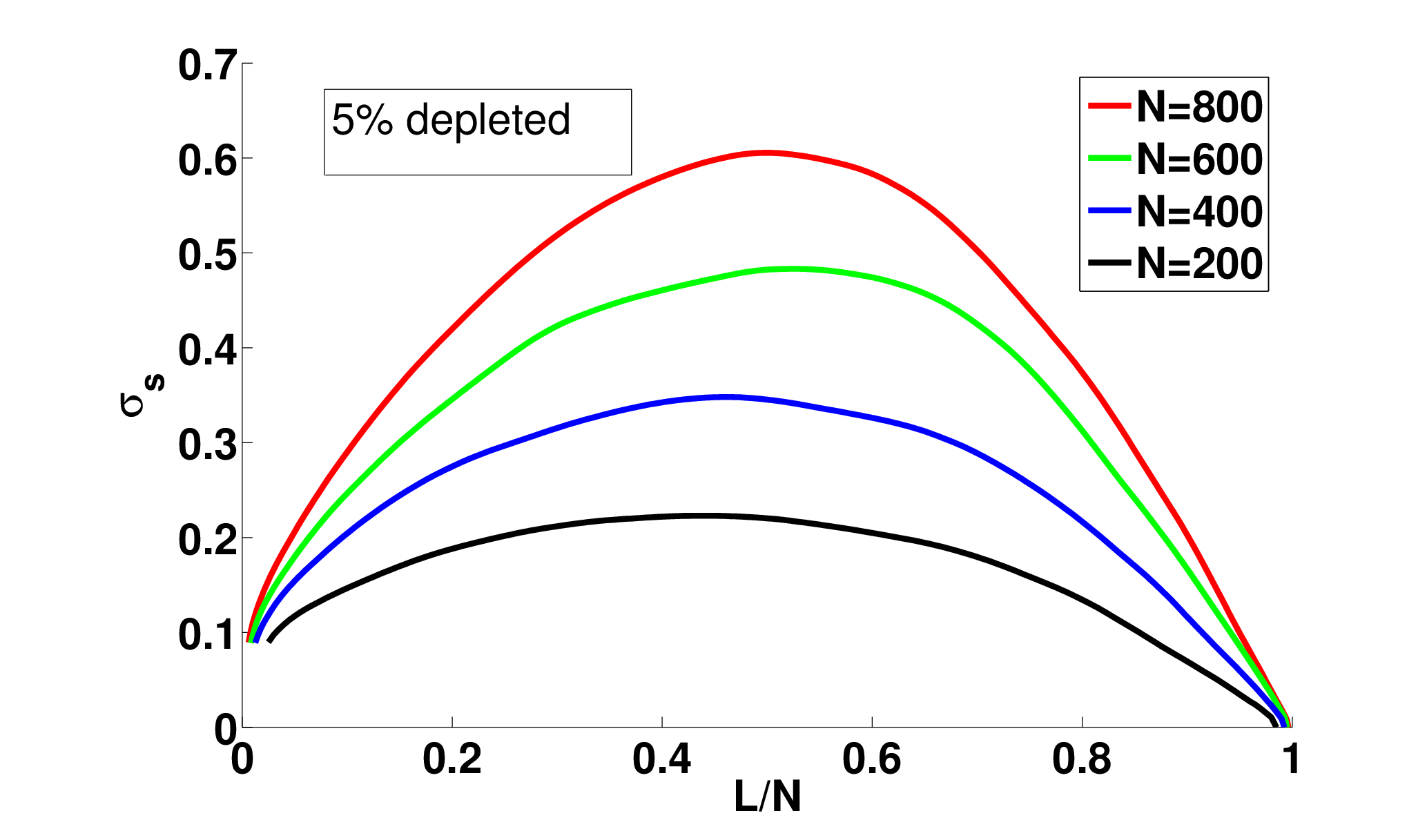}
\caption{\label{fig:sigma_sdep5} (Color online) The rescaled
standard deviations for systems of various sizes,
$\sigma_s(N,L;x)$, with 5\% of levels depleted. Each line
represents the average for 500 systems of a particular size.}
\end{figure}

\begin{figure}
\includegraphics[width=4in]{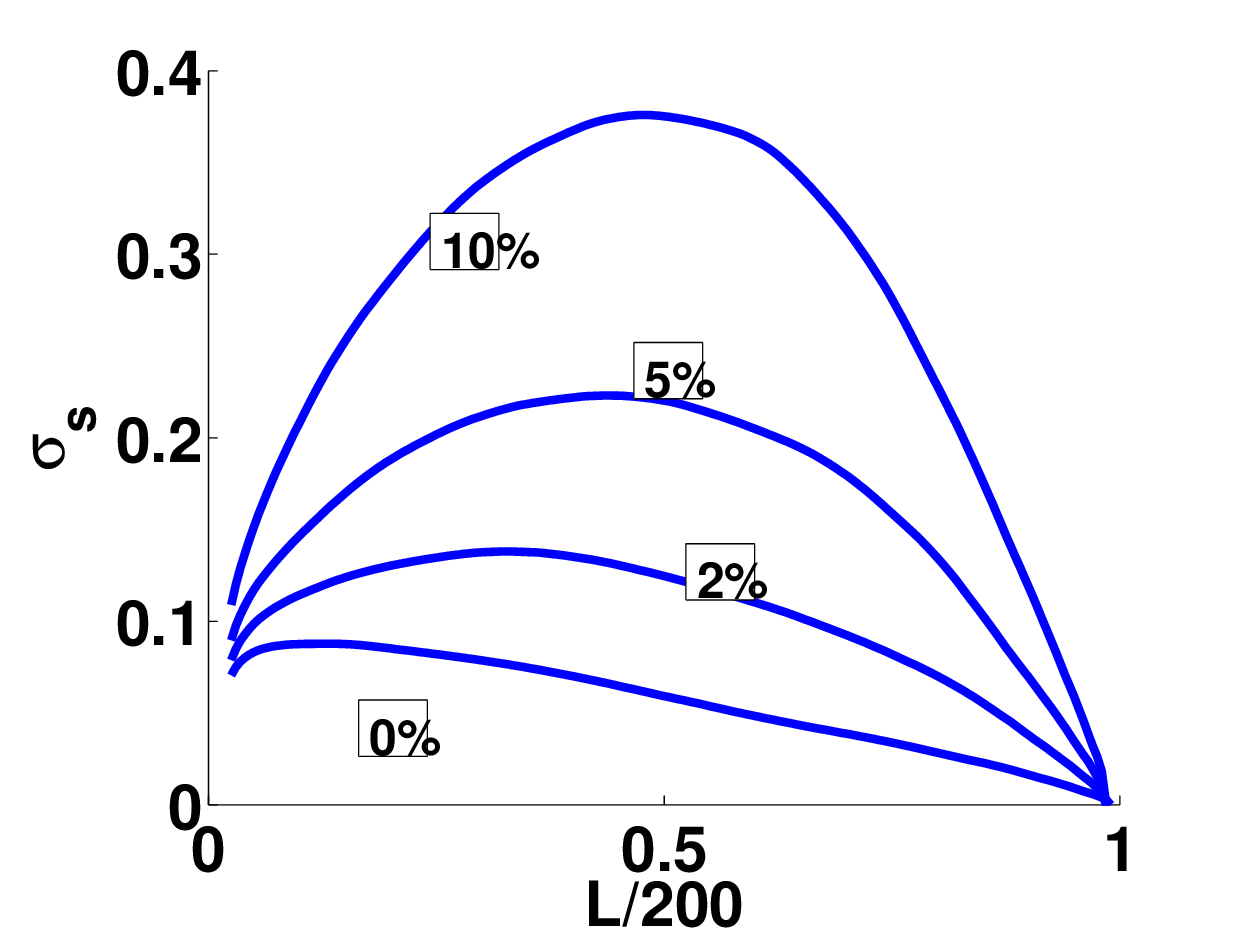}
\caption{\label{fig:sigma_sdep2N200} (Color online) The
rescaled standard deviations for $N=200$, $\sigma_s(N,L;x)$,
with varying degrees of depletion. Each line represents the
average of 500 systems.}
\end{figure}

It is interesting to see the behavior of $\sigma_s(N,L;x)$, the
spectral average of the variation of $\Delta_3(L;x)$. It has a
strong $N$- and $x$-dependence, and the simple rule $\sqrt{N-L}$ of
the $x=0$ case is lost. In Fig.~\ref{fig:sigma_sdep5} we show that
the $N$-dependence of $\sigma_s(N,L;x)$ for 5\% depletion peaks at
$L\approx N/2$. Strong correlation in overlapping windows $L$ levels
wide would reduce this number as $L$ increases, and in
Fig.~\ref{fig:sigma_sdep2N200} the results for $N=200$ and
$x=0\%,\,2\%,\,5\%,$ and $10\%$ show that the maximum moves to
higher values of $L$ as $x$ increases.

\section{\label{sec:ps}Level Spacing analysis}

The nearest level spacing distribution can be used to test for
missing levels, and mixing of different sequences. Here we will
follow the work of Agvaanluvsan {\sl et al.} \cite{agv}, where the
maximum likelihood method is used to find the fraction of missing
levels in a sequence. We will test the method on GOE spectra
depleted by hand. Single sequence neutron resonance data will then
be analyzed, and the results compared with those of a $\Delta_3$
analysis.

\begin{figure}
\includegraphics[width=4in]{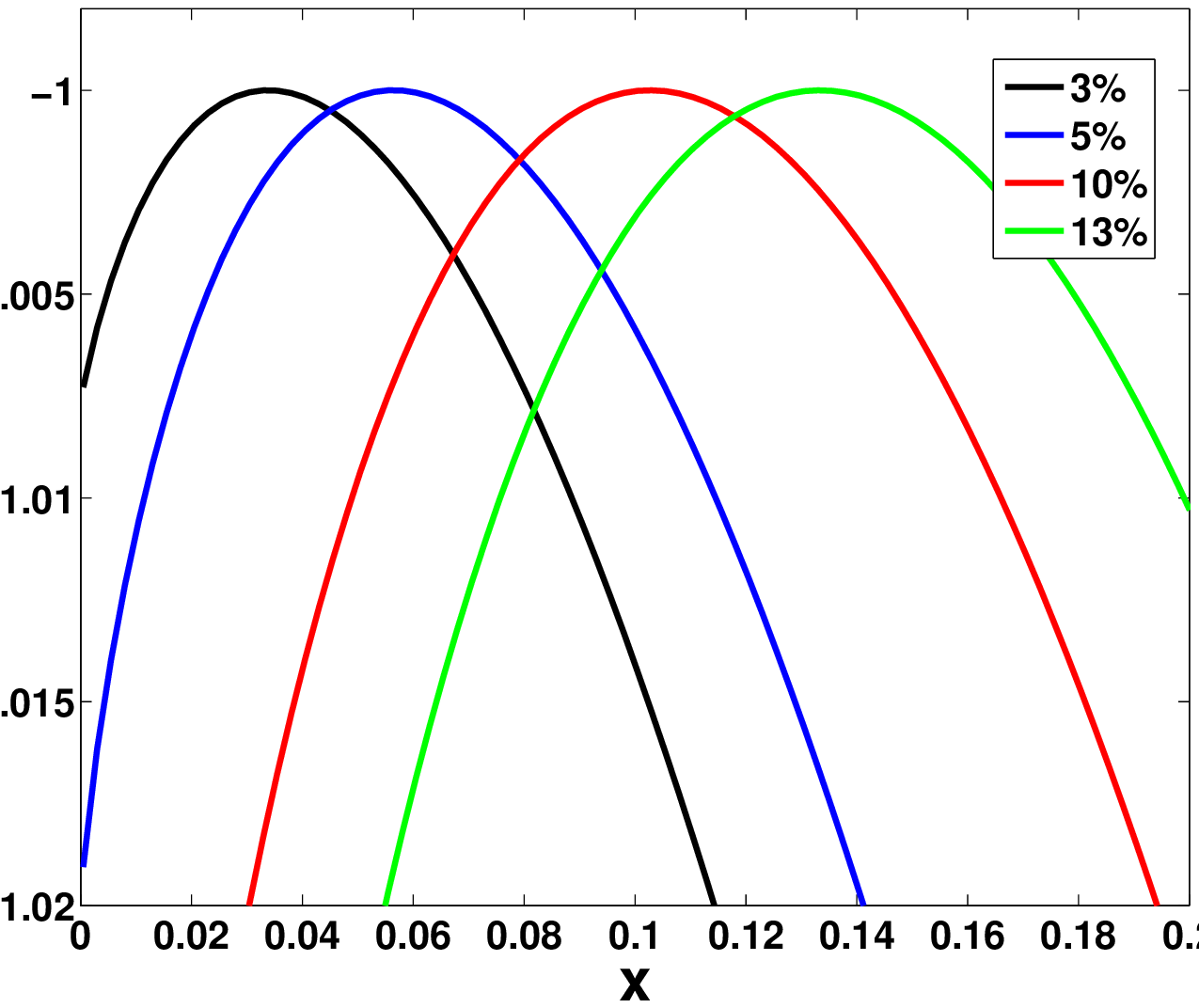}
\caption{\label{fig:mlm} (Color online) The maximum
likelihood method (MLM) is tested on a superposition of 200
GOE spectra, each of length $N=250$. A fraction $x$ of each
spectrum was randomly removed, and the MLM was used to
recover the number. Here are the results. The MLM slightly
overestimates the fraction missing. The natural log of the
likelihood function, $\ln({\mathcal L})$, is plotted against
$x$.}
\end{figure}

The level spacing distribution for a complete GOE spectrum is given
by the Wigner surmise,
\begin{equation}
P(s)=\frac{\pi}{2}s e^{-\pi s^2/4},
\end{equation}
where $s=S/D$, $S$ being the spacing between adjacent levels,
and $D$ is the average spacing. We have unfolded all spectra
involved, so $D=1$. If a level is missing, then two nearest
neighbor spacings are unobserved, while one  next-to-nearest
spacing is included as a nearest level spacing, when it
should not be. Furthermore, if $1-x$ is the fraction of the
spectrum that is observed, then $D_{{\rm obs}}$, the
experimental value for the average spacing, is related to the
true value by $D=(1-x)D_{{\rm obs}}$. Agvaanluvsan {\sl et
al.} show that
\begin{equation}
P(s)=\sum_{k=0}^{\infty} (1-x)x^kP(k;s),
\end{equation}
where $P(k;s)$ is the distribution function for the $k^{th}$ nearest
neighbor spacing, $E_{k+i}-E_i$; for $k=0$ this reduces to the
Wigner distribution, $P(0;s)=P(s)$.

The maximum likelihood method will be used to find the best
value for $x$ that maximizes the likelihood function
${\mathcal L}=\prod_{i}P(s_i)$; the product is over all the
observed spacings. In practice it is easier to maximize
$\ln({\mathcal L})=\sum_{i}\ln P(s_i)$. The functions
$P(k;s)$ are complicated to derive. For $k\neq 0$, the
functions were fitted to the empirical distributions from the
superposition of 5000 GOE spectra, each of length $N=5000$.
The procedure was tested on 200 GOE spectra, each of length
$N=250$, with 3\%, 5\%, 10\% and 13\% of levels randomly
removed. The results are shown in Fig.~\ref{fig:mlm}. The
method systematically overestimated $x$ by about 0.5\%.

\begin{table}
\caption{\label{tab:mlmerror}The errors in $x$, the fraction
of missed levels, using the maximum likelihood method. The
tests were run for 1500 depleted spectra of size $N=100$ and
300. The mean value $\overline{x}$, and the standard
deviation, $\sigma$ are given. }
\begin{ruledtabular}
\begin{tabular}{llllll}
$x$ & $\overline{x} \,(N=100)$ & $\sigma \,(N=100)$ & $\overline{x} \,(N=300)$ & $\sigma \,(N=300)$ \\
\hline
    2.91\%   &   3.32\%   &    2.42\%   &   3.44\%  &     1.49\% \\
    4.76\%  &     5.02\%  &     2.56\% &      5.29\%  &     1.43\% \\
    8.26\% &      8.45\%  &     2.48\%  &     8.74\%  &     1.32\% \\
    9.91\%  &     10.09\%   &   2.33\%   &   10.32\%  &     1.18\% \\
\end{tabular}
\end{ruledtabular}
\end{table}

In order to get an estimate of the error in $x$ from this method,
the procedure of the previous section was repeated, and the results
are shown in Table \ref{tab:mlmerror}. The agreement with the
$\Delta_3(L)$ method is encouraging. The $\Delta_3(L)$ method seems
to be slightly more accurate, but the uncertainty in $x$ is slightly
smaller for the MLM.

\begin{figure}
\includegraphics[width=3.5in]{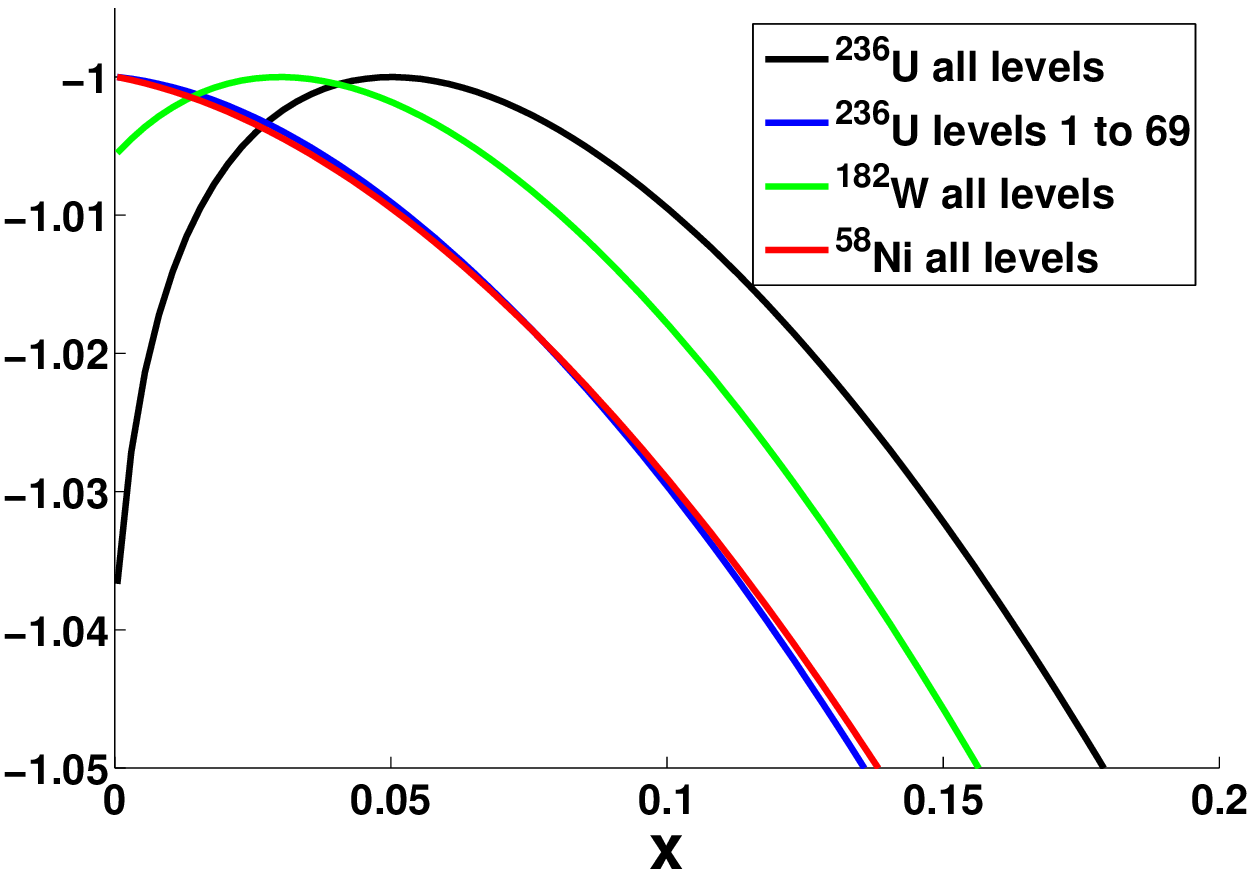}
\caption{\label{fig:mlmUNW} (Color online) MLM results are
shown for some representative data sets. The indication is
that the $^{236}\text{U}$ data is incomplete, with 5\% of
levels missing, but the subset of the data consisting of the
first 69 levels looks complete. There appears to be 3\% of
levels missing from the  $^{182}\text{W}$ data, with 0\%
missed from the $^{58}\text{Ni}$ data.}
\end{figure}

Next the procedure was applied to the real neutron resonance
data. The data sets are described in the next section. In
Fig.~\ref{fig:mlmUNW}  we see some typical results of the MLM
analysis. For the $^{236}\text{U}$ data, the full set of 81
levels looked incomplete, to the tune of about 5\%, while the
first 69 levels looked complete. The $^{182}\text{W}$ data
looked like there were 3\% of the 68 levels missing, while
the 63 levels in the $^{58}\text{Ni}$ data looked like a
complete set.

\section{\label{sec:data}Neutron resonance data}

The  neutron resonance data for a wide range of isotopes are now
widely available \footnote{See, for example, the Los Alamos National
Laboratory website http://t2.lanl.gov/cgi-bin/nuclides/endind.}.
Data sets with spin-0 target nuclei were chosen for analysis. They
afforded the simplicity of not having the fractional densities $f_1$
and $f_2$ of a superposition of two independent spectra, as all
levels have spin 1/2. There is a conventional assumption here that
the capture is dominated by $s$-wave neutrons, so no higher angular
momentum resonances intrude. Eleven isotopes were analyzed, and the
data sets are described in Table~\ref{tab:data}.

\begin{figure}
\includegraphics[width=3.5in]{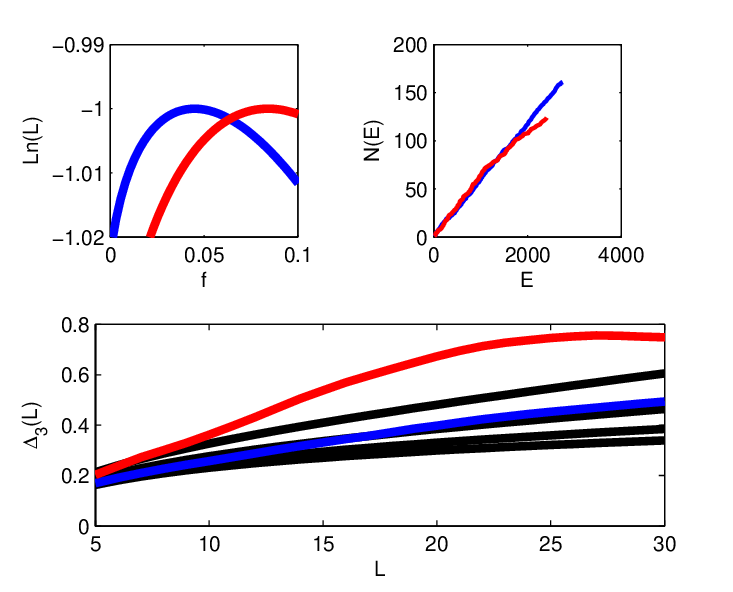}
\caption{\label{fig:Gd}(Color online) $\Delta_3(L)$ vs. $L$
for the experimental data. The black lines are the GOE values
for 0\%, 2\%, 5\%, and 10\%, starting from the lowest curve.
The $^{154}\text{Gd}$ data is in blue, and the
$^{152}\text{Gd}$ data is in red.}
\end{figure}

The cumulative level number gives the first indication of the purity
of the data. Kinks in ${\cal N}(E)$ leading to smaller slopes would
suggest a section of data where levels were missing. Using this as a
guideline, some data sets were split into subsets. For example, in
Fig.~\ref{fig:Gd} there is a kink in ${\cal N}(E)$ at the
$70^{\text{th}}$ level, so we analyzed the subset of levels
$E_1\rightarrow E_{70}$. The percentage of missing levels was
estimated with the maximum likelihood method. $\Delta_3(L)$ was then
calculated for each set, and compared with the MLM results. In
Fig.~\ref{fig:Gd} the three quantities are shown for two
representative isotopes, $^{154}\text{Gd}$ (blue), and
$^{152}\text{Gd}$ (red). The results are summarized in
Table~\ref{tab:data}. The agreement between the two methods is
promising.

\begin{table}
\caption{\label{tab:data}The results for $x$, the percent of
missing levels in the data. }
\begin{ruledtabular}
\begin{tabular}{llllll}
Isotope & MLM & $\Delta_3(L)$ & N (\# levels) & subset \\
\hline
$^{50}\text{Cr}$ & 2\% & 4\%   &   64 & $1 \rightarrow 36$  \\
$^{50}\text{Cr}$ & 0\% & 8\%   &   64 & $37 \rightarrow 64$  \\
$^{54}\text{Fe}$ & 3\% & 0\%  &   63 &$1 \rightarrow 50$ \\
$^{58}\text{N}$ & 0\% & 0\%  &    63 & All \\
$^{152}\text{Sm}$ & 3\% & 2\%  &     91 &$1 \rightarrow 70$ \\
$^{152}\text{Gd}$ & 8\% & 12\%  &   128&$1 \rightarrow 70$ \\
$^{154}\text{Gd}$ & 4\% & 2\%  &    161 &All \\
$^{158}\text{Gd}$ & 11\% & 7\%  &    93 &All \\
$^{158}\text{Gd}$ & 0\% & 1\%  &    93 &$1 \rightarrow 60$ \\
$^{158}\text{Gd}$ & 12\% & 15\%  &   93 &$61 \rightarrow 93$ \\
$^{182}\text{W}$ & 3\% & 2\%  &    68 &All \\
$^{234}\text{U}$ & 6\% & 4\%    &   118 &$1 \rightarrow 75$ \\
$^{234}\text{U}$ & 7\% & 3\%   &   118 &$76 \rightarrow 118$ \\
$^{234}\text{U}$ & 9\% & 8\%   &   118 & All\\
$^{236}\text{U}$ & 0\% & 1\%  &    81 &$1 \rightarrow 69$ \\
$^{236}\text{U}$ & 5\% & 8\%   &   81 & All\\
$^{240}\text{Pu}$ & 0\% & 3\%   &   267 & $1 \rightarrow 100$  \\
$^{240}\text{Pu}$ & 8\% & 12\%   &  267 & $171 \rightarrow 267$  \\
$^{242}\text{Pu}$ & 11\% & 13\%  &    67 & All  \\
\end{tabular}
\end{ruledtabular}
\end{table}

There is a range of behaviors exhibited by $\Delta_3(L)$. It can be
flatter than the RMT value, which suggests an artificially rigid
spectrum. This type of behavior is typical when the unfolding
procedure employs too specific a function for ${\cal N}(E)$, and all
the fluctuations are washed out, leaving the unfolded spectrum too
rigid. It can grow too rapidly, which could mean that there are more
independent spectra present than thought. This would happen if, for
example, $p$-wave neutrons were contributing to the data.

In the presence of sequences of energy levels with different spin
labels, we would have a superposition of independent spectra. This
is the case, for example, when $s$-wave neutrons are incident on a
target nucleus with spin $j_{{\rm tar}}$, so that the resulting
resonances have spin $j_\pm=j_{{\rm tar}}\pm\frac{1}{2}$. The level
density is relatively constant over the small energy range of the
data, and one would naively expect that the subspectra have
fractional densities $f_1=\frac{2 j_{-}+1}{(2 j_{-}+1)+(2 j_{+}+1)}$
and $f_2=1-f_1$. For $m$ independent spectra, with fractional
densities $f_1\,,f_2\dots f_m$, and letting $\Delta_{3m}(L)$ be the
spectral rigidity of the $m^{\rm{th}}$ sub-spectrum, we have
$\Delta_3(L)=\sum_{i=1}^m \Delta_{3m}(f_i L)$.

\begin{figure}
\includegraphics[width=3.5in]{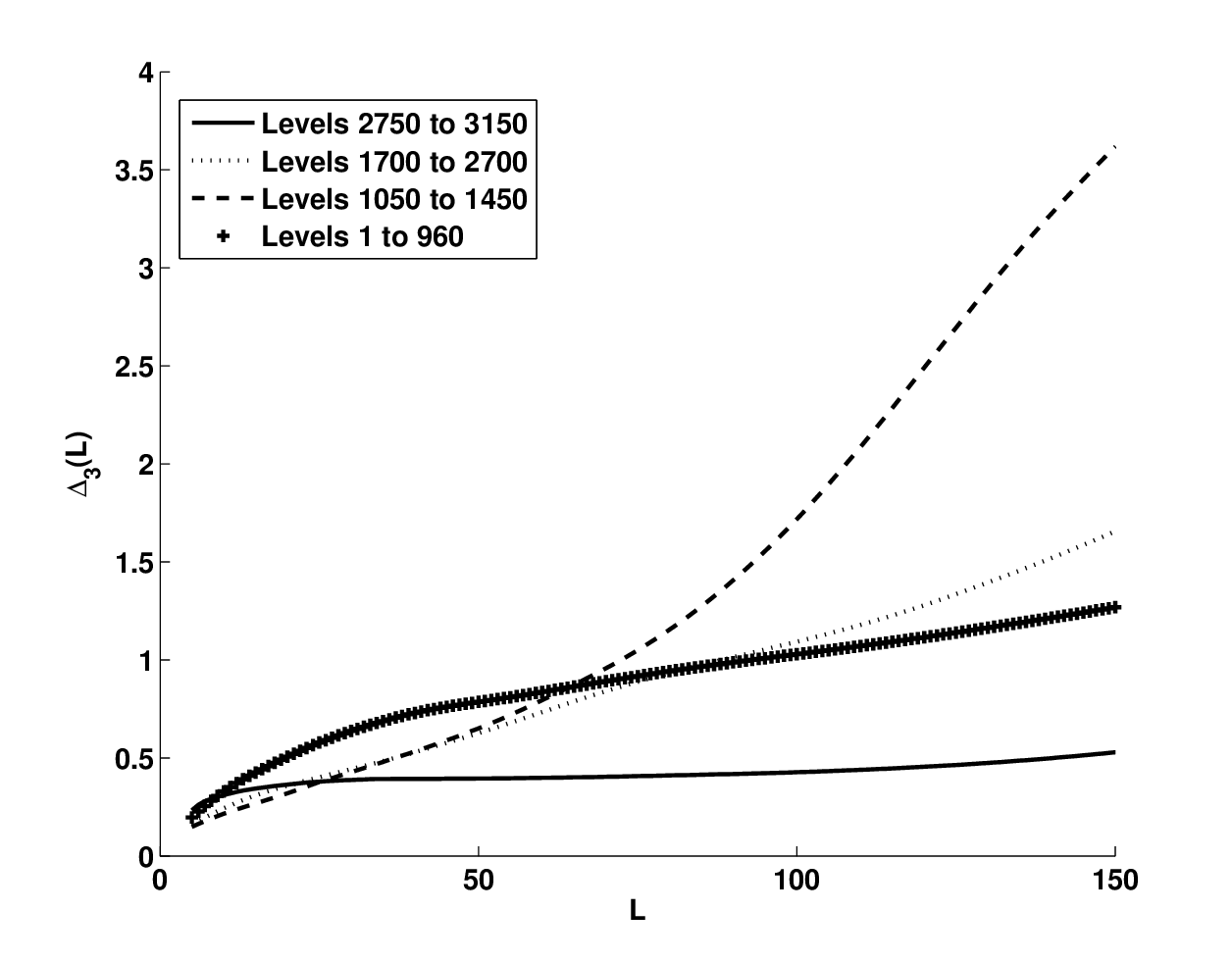}
\caption{\label{fig:u235all} $\Delta_3(L)$ vs. $L$ is plotted
for the four subsets of the $^{235}\text{U}$ data.The plot
suggests that the middle sections of the data are less pure
than the first set, but the highest energy set looks
artificially rigid, like a harmonic oscillator spectrum.}
\end{figure}

A preliminary analysis of systems of mixed independent spectra  was
done on the $^{235}\text{U}$ data \cite{leal97}. The $s$-resonances
have spins $j=3$ and 4, making the data a superposition of two
independent spectra, with $f_1=0.4325$. This data set exhibited all
the behaviors described above. This data set, the biggest by far,
was split into four sections corresponded to linear regions on the
${\cal N}(E)$ curve. The data was unfolded by fitting sections of
${\cal N}(E)$ to a straight line.  The first section had 950 levels
and the energy range from 0 to 510 eV. It exhibited excellent
agreement with the GOE result for mixed ($f_1=0.4$) spectra with 4\%
depletion. The next sets were levels 1050 to 1450, with range $578
~\text{eV}\rightarrow 920 ~\text{eV}$, 1700 to 2700 with range $1118
~\text{eV}\rightarrow 1995 ~\text{eV}$, and the fourth set had
levels 2750 to 3150, with range $2022 ~\text{eV}\rightarrow 2240
~\text{eV}$. The results are summarized in Fig.~\ref{fig:u235all}.
Although we do not have a MLM comparison for the case of two
independent spectra, the $\Delta_3(L)$ analysis suggests that 4\% of
the levels were missing in the first 960 levels.

\begin{figure}
\includegraphics[width= 3.5in]{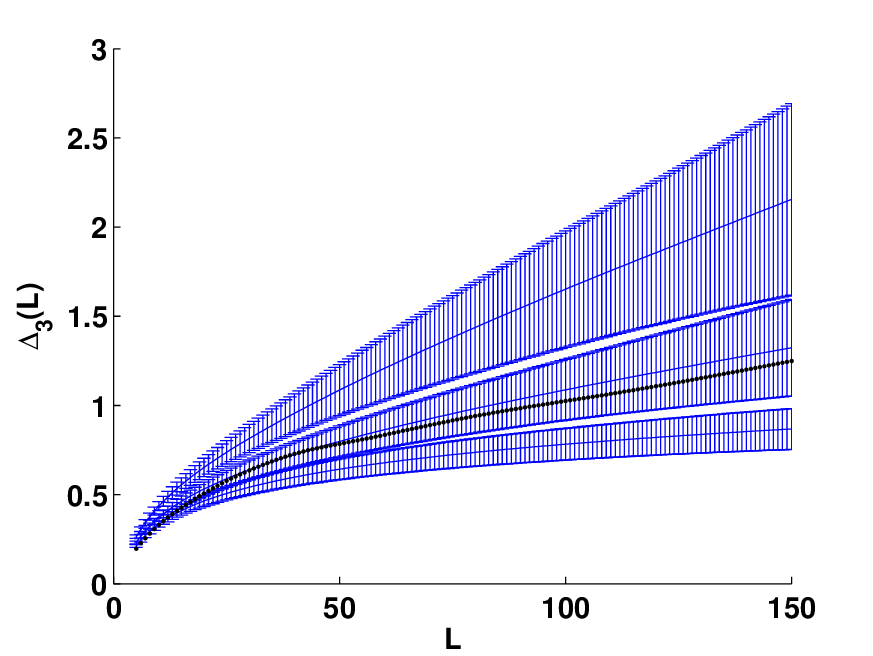}
\caption{\label{fig:u235low} (Color online) $\Delta_3(L)$ vs.
$L$ is plotted for the first 960 levels of the
$^{235}\text{U}$ data. The uncertainties are empirical for
mixed ($f_1=0.44$) GOE spectra, with 0\%, 4\%, and 10\$
depleted. The data is consistent with there being 4\% of
levels missing.}
\end{figure}

\section{Conclusion}

We have reexamined the possibility of using the Dyson-Mehta
$\Delta_3(L)$ statistic as a diagnostic for the spectra of quantum
systems in the chaotic regime, where the fluctuations can be modeled
by Random Matrix Theory. Originally it was regarded as a tool of
much more limited resolution, due to the large variance
\cite{dyson}. This is not the full picture, and the definition of
the statistic and the original results are reinterpreted. An
examination of the ergodicity of the statistic lead to the  spectral
properties being separated from ensemble properties. We see that it
is inappropriate to use the large and constant variance of the
statistic as an uncertainty. Uncertainties in $\Delta_3(L)$ were
found empirically for pure and depleted GOE spectra. In doing this,
the Poisson estimate of Brody {\sl et al.} was verified. The
$\Delta_3(L)$ statistic was used to determine the percentage of
missing levels in neutron resonance data. The results were compared
with the maximum likelihood method of Agvaanluvsan et al \cite{agv},
and the agreement was good. The method was applied to
$^{235}\text{U}$ data with  curious results.  Various sections of
the data displayed different behavior, raising questions about the
purity of the data, the accuracy of the angular momentum assignment,
and the appropriateness of the modeling the data with RMT. The
behavior of $\sigma_s$, the spectral spread of $\Delta^i_3(L)$, was
examined, and new properties were described which need a more
detailed study.
 It is hoped that the $\Delta_3(L)$ statistic
will be considered as a useful diagnostic in the RMT arsenal for
data analysis.

\begin{acknowledgments}

We wish to acknowledge the support of the Office of Research
Services of the University of Scranton, NSF grant PHY-0555366, and
the National Superconducting Cyclotron Laboratory at the Michigan
State University. We thank A. Volya , M. Moelter and F. Izrailev for
constructive discussions.

\end{acknowledgments}

\bibliography{d3neutronresPRC}

\end{document}